\PassOptionsToPackage{hyphens}{url}
\documentclass[conference]{IEEEtran}
\usepackage{amsmath,amssymb,amsfonts}
\usepackage{algorithmic}
\usepackage{graphicx}
\usepackage{textcomp}
\usepackage{xcolor}
\usepackage{url}
\usepackage{subcaption}
\usepackage{float}
\usepackage{pgf} 
\usepackage{svg} 
\usepackage{epsfig}
\usepackage[numbers,sort]{natbib}
\usepackage{makecell}
\usepackage{pgfplots} 
\usepackage{xspace}
\usepackage{booktabs}
\usepackage{enumitem}
\usepackage{tabularx}
\usepackage{comment} 
\usepackage{amsbsy} 
\usepackage{verbatim} 
\usepackage{pifont}
\usepackage{tablefootnote}
\usepackage{listings}
\usepackage{nopageno}
\usepackage{siunitx}
\usepackage[colorlinks, citecolor=darkgreen]{hyperref}
\usepackage[capitalise,nameinlink,noabbrev]{cleveref}
\def\BibTeX{{\rm B\kern-.05em{\sc i\kern-.025em b}\kern-.08em
    T\kern-.1667em\lower.7ex\hbox{E}\kern-.125emX}}

\newif\ifNotes\Notesfalse

\definecolor{darkgreen}{rgb}{0,0.65,0}
\definecolor{darkviolet}{HTML}{9400D3}
\definecolor{wildwatermelon}{HTML}{FF43A4}

\newcommand{\parhead}[1]{\vspace{1pt plus 1pt minus 1pt}\par\noindent\textbf{#1}\hspace{.4em plus .2em minus .2em}}

\newcommand{\instr}[1]{{\fontfamily{qcr}\selectfont #1}\xspace}
\creflabelformat{equation}{#2\textup{#1}#3}

\begin{document}
\title{Kraken: Higher-order EM Side-Channel Attacks on DNNs in Near and Far Field}

%\author{\IEEEauthorblockN{Michael Shell}
%	\IEEEauthorblockA{Georgia Institute of Technology\\
%		someemail@somedomain.com}
%	\and
%	\IEEEauthorblockN{Homer Simpson}
%	\IEEEauthorblockA{Twentieth Century Fox\\
%		homer@thesimpsons.com}
%	\and
%	\IEEEauthorblockN{James Kirk\\ and Montgomery Scott}
%	\IEEEauthorblockA{Starfleet Academy\\
%		someemail@somedomain.com}}
	

\author{\IEEEauthorblockN{Peter Horvath\IEEEauthorrefmark{1},
Ilia Shumailov\IEEEauthorrefmark{2},
Lukasz Chmielewski\IEEEauthorrefmark{3},
Lejla Batina\IEEEauthorrefmark{1}
Yuval Yarom\IEEEauthorrefmark{4}
}
\IEEEauthorblockA{\IEEEauthorrefmark{1} Radboud University, Nijmegen, Netherlands}
\IEEEauthorblockA{\IEEEauthorrefmark{2} AI Sequrity Company, London, United Kingdom}
\IEEEauthorblockA{\IEEEauthorrefmark{3} Masaryk University, Brno, Czech Republic}
\IEEEauthorblockA{\IEEEauthorrefmark{4} Ruhr University, Bochum, Germany}
}


\maketitle
\begin{abstract}
    The multi-million dollar investment required for modern machine learning (ML) has made large ML models a prime target for theft. In response, the field of model stealing has emerged. Attacks based on physical side-channel information 
    have shown that DNN model extraction is feasible, even on CUDA Cores in a GPU.    
    For the first time, our work demonstrates parameter extraction on the specialized GPU's Tensor Core units, most commonly used GPU units nowadays due to their superior performance, via near-field physical side-channel attacks. 
    Previous work targeted only the general-purpose CUDA Cores in the GPU, the functional units that have been part of the GPU since its inception. 
    Our method is tailored to the GPU architecture to accurately estimate energy consumption and derive efficient attacks via Correlation Power Analysis (CPA). Furthermore, we provide an exploratory analysis of hyperparameter and weight leakage from LLMs in far field and demonstrate that the GPU's electromagnetic radiation leaks even 100\,cm away through a glass obstacle.
\end{abstract}

\begin{IEEEkeywords}
Side-Channel Analysis (SCA), GPU, Large Language Models, Electromagnetic Radiation, Model Stealing.
\end{IEEEkeywords}

\section{Introduction}\label{sec::intro}
Deep neural networks (DNNs) are deployed in ever more applications, and this trend does not appear to be slowing down since the introduction of ChatGPT, a Transformer-based~\cite{NIPS2017_3f5ee243} architecture that prompted unprecedented investments in this technology. DNNs can require up to tens of millions of dollars to train, and the models' secrets are Intellectual Property (IP). Therefore, companies aim to protect their models; otherwise, their competitive advantage suffers.
An adversary who might otherwise be unable or unwilling to train their own model might be tempted to steal these models. In addition to the enormous costs of just training these models, many of the datasets used to train them are proprietary, making it difficult for anyone to develop their own models. 

Although state-of-the-art models can have billions to trillions of weights, making extraction hard, in many practical settings, an adversary does not necessarily need to extract all the weights. In a widely used scenario where an open source base model is fine-tuned using Low-Rank Adaptation (LoRA)~\cite{lora_weights}, only a small subset of weights is changed. Therefore, the adversary only has to extract this subset of weights.

There are different ways to steal model weights, depending on the capabilities of the adversary.
Most of the investigated methods rely on the unhindered attacker's access to the API and are affected by, for example, rate limiting, quantisation~\citep{carlini2020cryptanalyticextractionneuralnetwork}, or adversarially constructed noise~\citep{kariyappa2020adaptive}. 
Another threat is physical side-channel attacks on deep learning models, which have been successfully demonstrated on different platforms such as microcontrollers (MCUs)~\cite{batina2019csi, Joud2022:PracticalSCADNN}, FPGAs~\cite{Gongye2023:SCA-DPU}, and GPUs \cite{horvath2025barracuda} using power or EM side-channel information. These attacks can target different components of DNNs, such as the architecture, weights, or inputs. Although the leakage of Convolutional Neural Networks (CNNs) architectures has been exploited in the far-field of the electromagnetic region~\cite{liang2022clairvoyance}, weight extraction, especially of LLMs, has remained an uncharted territory. In general, prior work builds on the principles of Differential Power Analysis~\cite{kocher1999differential}, targeting weight-dependent intermediate values to demonstrate weight extraction. However, these attacks do not account for the parallel nature of GPUs and other hardware DNN accelerators.
More broadly, parallel architectures require refined leakage models to more accurately characterize the energy consumption of the target. Therefore, to the best of our knowledge, we are the first to consider the GPU architecture and its implementations in detail to derive \emph{warp-level} leakage models that significantly improve over state-of-the-art results by considering the energy consumption of all threads in a warp.%\lukasz{please double check this sentence, is it OK now?}

However, there are other aspects of DNNs that are not yet taken into account for weight extraction, which, to the best of our knowledge, allows us to introduce \emph{higher-order} side-channel attacks on DNNs. It is important to recognize that for a typical DNN executing on any platform, there are multiple intermediate values that depend on the same weight. For instance, in Convolutional Neural Networks (CNN), each kernel is convolved with many parts of the input. For LLMs and attention layers, each column of the projection matrices computes multiple dot products with embeddings, where the number of dot products depends on the number of input tokens. In both examples, many dot products depend on the same weights and thus provide more information to an adversary. In this paper, we demonstrate how to combine information from multiple intermediate values, thereby significantly speeding up our 
%side-channel 
attack. 
%\lukasz{does it sound ok?}\peter{yes, thanks}
Furthermore, to the best of our knowledge, we present the first proof-of-concept far-field EM side-channel attack on a 4\,090 RTX GPU for LLMs, showing that leakage can be observed in the far field as well.

To summarize, our contributions are as follows.
\begin{enumerate}
    \item We provide the first GPU-specific \emph{warp-level} approach and the first platform-agnostic \emph{higher-order} attack, both of which are more efficient than the state of the art. The warp-level model more accurately characterizes GPU energy consumption, while the higher-order attack further enhances the efficiency of near-field attacks.
    \item To the best of our knowledge, we also present the first EM side-channel-based analysis of an LLM in the far-field of the electromagnetic region.
    We systematically investigate the leakage of different information (e.g., the number of generated tokens) of LLMs in the far-field region. We also analyze the weight-related leakage of LLMs in the far field using a proof-of-concept attack. 
    \item We discuss how quantization of the weights leaks information to an attacker, even before mounting any side-channel attacks.
\end{enumerate} 

\parhead{Organization} The paper's outline is as follows. First, we give the necessary background in~\cref{sec::bg} and mention relevant previous works in~\cref{sec::rw}. Next, we provide an overview of the attacks in~\cref{sec::overview}. Our near and far field attacks are described in~\cref{sec::nfa} and~\cref{sec::ffa}, respectively, followed by a discussion in~\cref{sec::disc}. Finally, the conclusions are given in~\cref{sec::conc}.

\section{Background}\label{sec::bg}
\subsection{Neural Networks}
LLMs are autoregressive neural networks that can take as input various modalities, such as text and image, and produce outputs conditioned on those inputs. The attention mechanism, one of the main building blocks of LLMs, provides input-dependent weights: $\mathit{Attention(\mathbf{X}) = \mathbf{X}\cdot\mathbf{W(X)}}$.
There are three matrices, $\mathbf{Q} = \mathbf{W}_q \cdot \mathbf{X}   $ (query), $\mathbf{K} = \mathbf{W}_k \cdot \mathbf{X}   $ (key), $\mathbf{V} = \mathbf{W}_v \cdot \mathbf{X} $ (value), derived from the input embeddings. The matrices $\mathbf{W}_q, \mathbf{W}_k, \mathbf{W}_v$ contain the fixed weights of the model and are the primary interest of an adversary. We assume the inputs, $\mathbf{X}$, to be known to the attacker. 
%lukasz{This is only about LLM and not NNs in general. Do we need more here?}

\subsection{Side-Channel Analysis}

Physical Side-Channel Analysis (SCA) exploits unintended information leakage of electronic devices, based on physical properties, to extract secret information such as cryptographic keys~\cite{kocher1996timing, kocher1999differential}.
The crucial insight of SCA is that the measurable physical properties, such as power consumption, depend on the operations executed and the data processed by the device.
There are multiple types of leakage that can be exploited, including power consumption, electromagnetic (EM) emanations, timing, optical, or sound. 
In this work, we exploit the EM side channel emanating from the die that contains the GPU on which a neural network is running, targeting the secret weights of the model with Correlation Power Analysis.\\

\noindent \textbf{Correlation Power Analysis}
In Correlation Power Analysis (CPA)~\cite{brier2004correlation}, Pearson's correlation coefficient is used as a side-channel distinguisher, i.e., to rank the key candidates.
By targeting an intermediate computation that depends on a secret and a known value, an attacker can build hypothetical power predictions by enumerating all possible candidates for the secret value and correlating them with the real power measurements. In SCA research on cryptographic implementations, these intermediate values are secret key-dependent values, such as the output of SBox of AES~\cite{daemen1999aes}, which depends on a single key (byte) and input byte. 
For neural networks, similarly, the known values are the inputs, while the secrets are the weights. The intermediate computations that depend on these values are the partial sums in neural networks~\cite{horvath2024sok, horvath2025barracuda}. In DNN implementations, the data type can range from unsigned/signed integers to floating-point values. Additionally, the precision used for inference is diverse, ranging typically from 4 to 16 bits.
In terms of leakage models, the Hamming-Weight (HW) and the more general  Hamming-Distance (HD) models have proven to be effective to extract weights~\cite{batina2019csi, Gongye2023:SCA-DPU, horvath2025barracuda}.

\noindent \textbf{Electromagnetic radiation} Electromagnetic (EM) radiation is fundamental to our everyday lives, including applications like communications or health care, but unintended radiation can break the confidentiality of secret information.
Examples include breaking cryptographic implementations~\cite{QS01, kuhn1998soft, gandolfi2001electromagnetic}, reverse-engineering neural networks~\cite{batina2019csi, chmielewski2021reverse,horvath2025barracuda} and eavesdropping on display units~\cite{elibol2012realistic, hongxin2009recognition, liu2020screen}. 
Specifically, in the far field, AES implementations have also been targeted via EM side channel~\cite{ffaes, ffaes2}.
EM radiation is modeled as a wave of the electromagnetic field propagating through free space when charged particles accelerate. The relationships among charges, current, and changes in electric and magnetic fields are governed by Maxwell's equations. When a source, such as an antenna, emits radiation, we can distinguish between multiple regions of the electromagnetic field: reactive near-field, radiating near-field, and far-field. The boundaries between these regions depend on the dimensions of the radiating source as well as the wavelength of the radiation.

A well-known formula to establish the boundary $R$ between the near and far fields is $\mathit{R = \frac{2 \times D^{2}}{\lambda}}$~\cite{near-far-field} with $D$ the largest dimension of the radiating source and $\lambda$ the wavelength of the radiation.

\subsection{GPU Architecture}
GPUs are composed of Streaming Multiprocessors (SMs) that allow an application to execute hundreds of threads on individual SMs, using the Single-Instruction-Multiple-Threads (SIMT) model. When an application launches work on the GPU, the threads are distributed among the available SMs. In each SM, the scheduled threads are further divided into groups of 32 threads, called warps. Threads in a warp start execution together and are free to diverge individually, but efficient implementations aim to minimize thread divergence. For SCA, analyzing the power consumption of a thread~\cite{horvath2025barracuda} has proven to be a possible technique for extracting DNN weights with near-field access using CPA. In this work, one of our contributions is to study the GPU architecture in detail and derive more efficient leakage models based on all threads in a warp. In addition, we extend these techniques to LLMs and in the far-field region of the electromagnetic field.

\subsection{Model Stealing}

Model stealing research community has predominantly focused on query-based model APIs, evolving from early hard and soft-label extraction methods through distillation~\citep{jagielski2020highaccuracyhighfidelity} to more modern techniques that solve for model parameters directly~\citep{carlini2020cryptanalyticextractionneuralnetwork, foerster2024slowsignshighfidelitymodel,carlini2024polynomialtimecryptanalyticextraction}. More recent works also aim to scale it from classical small network extraction settings to modern large Transformer models~\citep{carlini2024stealingproductionlanguagemodel}. However, this entire class of attacks is fundamentally limited by its reliance on the API, which allows defenders to implement countermeasures like rate limiting and output obfuscation. In this work, we demonstrate a new district threat vector that infers model parameters directly from physical EM emanations, rather than focusing on API-based extraction. Fascinatingly and unexpectedly, we show that such attacks are possible in the far field and even through glass. 

\section{Related Work}\label{sec::rw}
Side-channel attacks aiming to extract DNN weights via DPA-based methods were introduced in the seminal work of Batina et al.~\cite{batina2019csi}.
Since then, different platforms, including FPGA and GPUs, have been shown to be vulnerable to DPA-based side-channel attacks by targeting MLP or CNN architectures~\cite{Dubey2020:Maskednet,Regazzoni2020:MLHardwareSecurity,Yli2021:ExtractionBNN,DBLP:conf/iscas/YoshidaKOSF20,Li2022:PowerAttacksDNN,Gongye2023:SCA-DPU,Joud2022:PracticalSCADNN, horvath2025barracuda}. For a more comprehensive comparison, we refer the reader to ~\cite{DBLP:conf/uss/HorvathL0B24}.

In the most recent related work on extracting the weights from GPU~\cite{horvath2025barracuda}, CUDA implementations that use the general-purpose CUDA Cores are attacked. In contrast, our work attacks implementations that use Tensor Core instructions. These instructions use hardware specifically designed to accelerate matrix multiplication. Implementations of LLMs heavily rely on the acceleration provided by Tensor Core units because the throughput by these units is much higher than that of the general-purpose CUDA cores. As we show in ~\cref{ssec::llama_impl}, commonly used models such as Llama 3 also use these units for acceleration. %\lukasz{Could we add something: ...matrix multiplication resulting in more efficient implementation and therefore Tenso Core instructions are broadly used in real-world applications nowadays... Maybe even add an example of such an application if you know any? Would that make sense?} \peter{Added a couple of sentences}
Essentially, there is greater parallelism with Tensor Core instructions, which results in more noise when we aim to reduce the energy consumption of a single thread. 
%is considered for energy consumption. 
This increased parallelism requires the \emph{warp-level} leakage model to reliably extract weights. Since newer GPUs and their implementations rely heavily on Tensor Cores, we consider this scenario more practical and closer to what an adversary might encounter in the real world. 
Furthermore, we introduce, for the first time, \emph{higher-order} attacks in the context of DNNs to significantly improve the convergence of weight extraction. 
Lastly, as a first study, we explore side-channel leakage of LLM hyperparameters and weights, and we do this exploration in the far field.
%\lukasz{does it sound better?}
%\lukasz{was there already SCA paper on LLMs in near field? also please double check this paragraph - I made some modifications. } \peter{Nope, no LLM attack in near field}

\section{Overview}\label{sec::overview}

\begin{figure*}
	\centering
	\includesvg[width=450pt, height= 800pt]{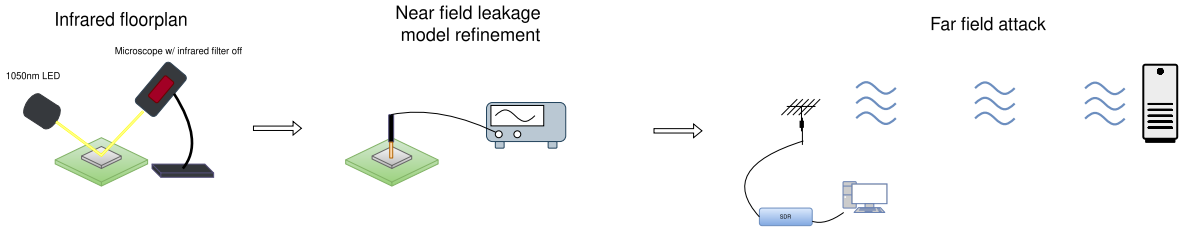}
	\caption{\label{fig::overview} High-level description of the attacks presented in the paper. First, a floorplan is made of the target chip to precisely identify the GPU SMs on the die. Second, near-field probing is used to test and validate the hypotheses of our leakage models, including warp-level and higher-order attacks. The approaches in the near field are validated on convolutional layers. Lastly, we provide an exploratory analysis of the leakage of different hyperparameters and the weights of an LLM in the far field using the derived leakage models in the near field.}
\end{figure*}

We illustrate our approach in \cref{fig::overview}. Our goal is to derive accurate leakage models for the GPU's energy consumption and extract weights via EM side-channel information. To that end, in \cref{sec::nfa} we start with analysis in near field by creating a floorplan of target die using infrared imaging in \cref{ssec::gpufloorplan}. The floorplan aids in targeting the components of the GPU that can leak information. Although it is possible to scan the whole die blindly to find exploitable points, this is unnecessary and requires enormous effort.
%on the die, 
Systems on a Chip (SoCs) contain many components, such as CPU cores, that are irrelevant to our attack, except that they generate additional noise that could hinder physical attacks in general.

Next, we aim to derive efficient leakage models for GPUs. 
We created two different models: one GPU-specific and a non-GPU-specific leakage model.
For the GPU-specific model, in \cref{ssec:timpl} we analyze the GPU implementations using hardware-accelerated General Matrix Multiply (GeMM) operations on the Tensor Cores of the GPU. Note that Tensor Core implementations were not analyzed in previous work~\cite{horvath2025barracuda}.
More specifically, we take into account how parallelism is implemented on the GPU with groups of threads called warps, hence we derive the \emph{warp-level} leakage model in \cref{ssec:timpl}. This model is a more accurate reflection of the energy consumption and provides much better results than the state-of-the-art~\cite{horvath2025barracuda}.

For the non-GPU specific leakage model, we provide the insight that, since every weight
interacts independently with different inputs, we can greatly enhance our attack by mounting \emph{higher-order} attacks introduced in ~\cref{ssec:holm}. In other words, there are multiple intermediate values that depend on the same weight, and this property of the neural network has not been exploited before in the SCA context. Furthermore, this method applies to other platforms and is not specific to GPUs but to DNNs.
Higher-order attacks are well-known in SCA as a means to defeat masking countermeasures in cryptographic implementations and to extract secret keys. However, in our case, the goal with the higher-order attack is to reduce the number of measurements needed, and the result of the attack is presented in ~\cref{ssec::wer}.

Finally, we present our far field attack on LLMs in ~\cref{sec::ffa}. First, we describe the threat model that applies to the adversary and their capabilities in ~\cref{ssec:threat}.
Next, we describe how the far field signal is modeled and the corresponding experimental setup to collect far field measurements in ~\cref{ssec::ffmodel} and ~\cref{ssec:exp}. The chosen LLM implementation is introduced in ~\cref{ssec::llama_impl}. 
Subsequently, we illustrate how the complexity can be reduced for the CPA attacks if chosen inputs are within an adversary's capabilities in ~\cref{ssec::cpa_comp}. We show our experiments regarding weight and input-related leakages in ~\cref{ssec::ileakage} and ~\cref{ssec::wleakage}. We show the impact of the number of generated tokens and the batch size on the signal in ~\cref{ssec::ltokenbatch}
We present the parameter extraction results for far-field in ~\cref{ssec::ff_cpa}.

In~\cref{sec::disc}, we provide additional details on the limitations of our work, on how quantization leaks information to an attacker and what countermeasures could be applied against the attacks presented in the paper.

Finally, we conclude the paper in ~\cref{sec::conc}.

\section{Near Field Attack}\label{sec::nfa}
\subsection{Motivation}
The computations in neural networks are dominated by matrix and pointwise operations that have to be parallelized to achieve sufficient performance in practice. From a side-channel perspective, in general, parallelization introduces noise in the measurements that makes an adversary's task harder. However, the amount of parallelization and introduced noise depends a lot on the target hardware architecture and the sophistication of an adversary's measurement setup. 

General-purpose GPU architectures manage large numbers of threads in small groups of 32~\cite{nv_warp, amd_hw} or 64~\cite{amd_hw}. Locality is crucial for performance, and the computations of each thread are not arbitrary.
For deep learning workloads and GPU implementations, this means that threads in a group work on the same slices of weights or inputs, or both. Additionally, threads generally execute in parallel physically; therefore, the sum of their individual energy consumption is a more accurate model of the overall energy consumption. Previous work used the energy consumption of a single thread, and in this work, we show that considering the energy consumption of all threads in a group that work on the same weights provides more efficient attacks. Furthermore, we show that considering threads across groups provides more efficient attacks than the state of the art. 
\subsection{GPU Die Floorplan}\label{ssec::gpufloorplan}

\begin{figure*}
	\centering
	\includesvg[width=0.8\linewidth]{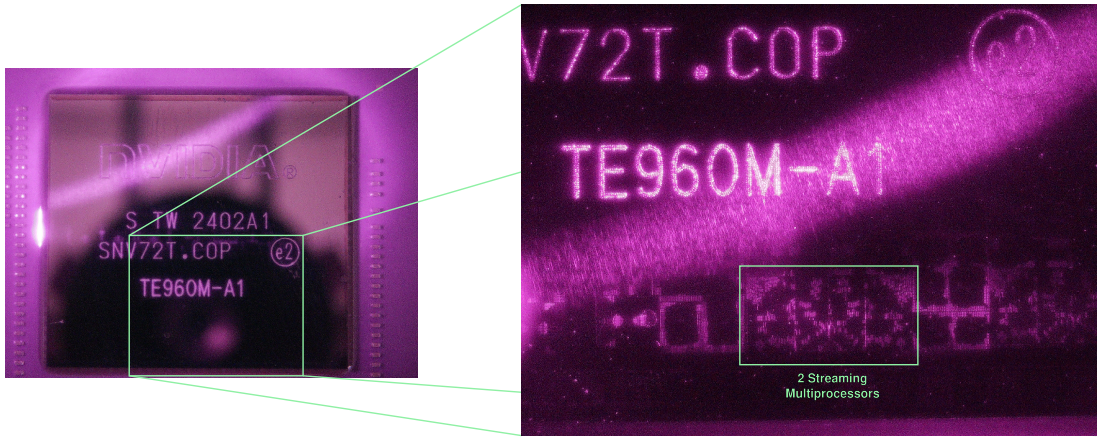}
	\caption{\label{fig::full_die} Image of the full die on the left and a partial infrared die shot on the right. The highlighted area with the rectangle on the right shows 2 Streaming Multiprocessors out of 16 of the GPU die.}
\end{figure*}

\begin{figure*}
	\centering
	\includesvg[width=0.8\linewidth]{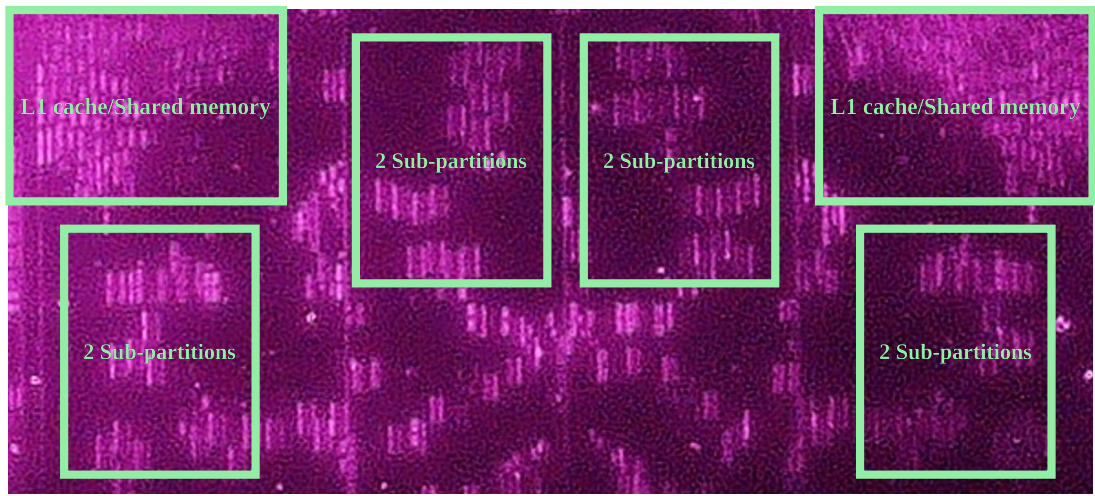}
	\caption{\label{fig::diesm_zoom} Zoomed in image of the two SMs with annotations. The partitioned L1 cache/Shared Memories are in the upper-left and upper-right corners. More importantly, each SM contains 4 sub-partitions, each with its own Tensor Core unit. We place the EM probe directly above or in the vicinity of one sub-partition.}
\end{figure*}

The Jetson Orin Nano uses flip-chip packaging without a heat spreader, meaning the silicon is directly exposed to the attacker. In general, the die in flip-chip packaging can be easily exposed without any invasive modification, such as chemical etching, that is required for other packaging types. Due to the transparency of silicon to infrared light, it is possible to inspect the structures of the die~\cite{huang2023infraredinsituirisinspection}. 
We show the full die and an infrared die shots of the Jetson Orin Nano GPU in \cref{fig::full_die} and \cref{fig::diesm_zoom}, respectively. The green rectangle highlights the structure of the GPU's two SMs.
There is a clear symmetry in the rectangle, indicating the presence of two similar hardware blocks. The whole die has 16 SMs overall, of which 8 are functional on this GPU. By running applications on the GPU, it is straightforward to determine the active SM's on the whole GPU die. However, the application should launch enough threads 
%and therefore warp 
to ensure all SMs receive warps to execute. %\lukasz{I think it is clearer now?}

\cref{fig::diesm_zoom} shows a zoomed-in image of the two SMs. Based on the visible structures, we can annotate the structures in the image, highlighting the 4 sub-partitions and the combined L1/Shared Memory of each SM.
Since the sub-partitions contain the register files and ALUs, it is reasonable to assume that the vicinity of the sub-partitions is where the EM probe should be placed.
Based on experiments, we can confirm that we can reduce the search space for placing the electromagnetic probe by just focusing on the areas at or near the sub-partitions, as these contain the register files where the registers are overwritten with the outputs of the Tensor Core instructions, i.e., sensitive intermediate values.
To assess whether a particular location leaks and which locations leak the most, we perform Test Vector Leakage Assessment (TVLA)~\cite{gilbert2011testing} in two ways: fixed vs.\ random input and fixed vs.\ random weight as in ~\cite{horvath2025barracuda}. TVLA is the most common way in practice to evaluate whether a target is leaking side-channel information.

\subsection{Trace Collection Setup and Preprocessing}
\begin{figure}
	\centering
	\includesvg[width=\linewidth]{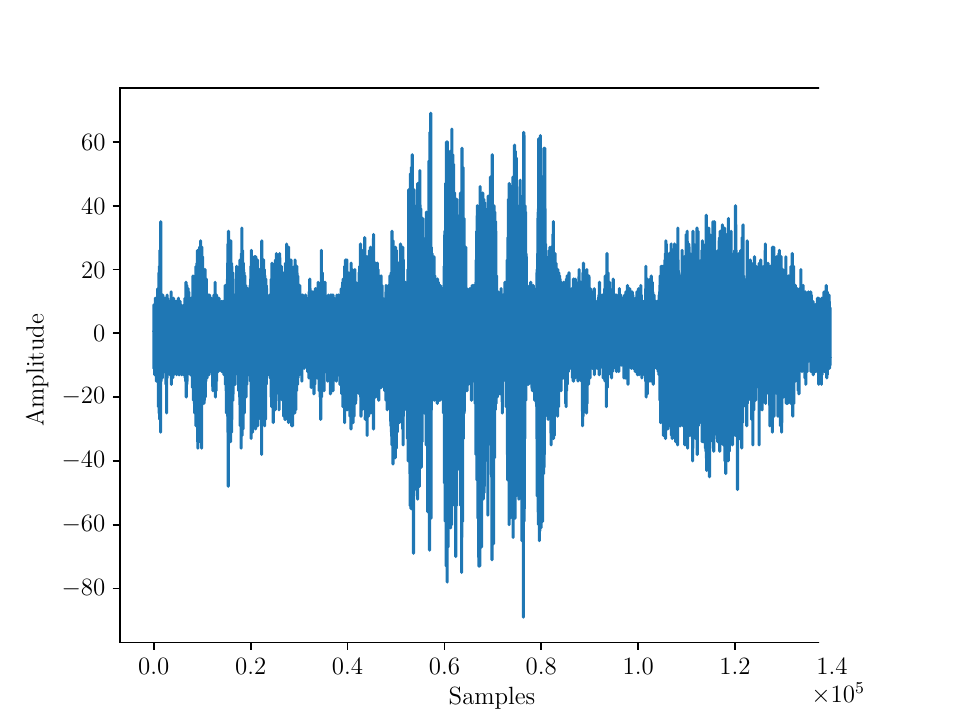}
	\caption{\label{fig::trace} Example of collected raw trace of a convolutional layer implemented on Tensor Cores. The vertical red lines highlight the Points-of-Interest (PoI) used for weight extraction.}
\end{figure}

For the measurements, we use the Langer RF-B 0.3\-3~\cite{langer_probe_rf} electromagnetic probe with the PA 303 amplifier and Lecroy Waverunner 8404-MS oscilloscope sampling at 20 GS/s. %\lukasz{it might be nice to mention the diameter of the probe}
Our target GPU is the Nvidia Jetson Orin Nano~\cite{orin_soc}. As opposed to the 625 MHz clock frequency in BarraCUDA~\cite{horvath2025barracuda}, we set the GPU cores clock frequency of the Jetson Orin Nano to 920 MHz. 
Similarly, our target neural network is a 2-layer CNN with 32 kernels in each layer of size $3\times3$ INT8 weights, using implementations from TensorRT~\cite{tensorrt}. However, our implementations use Tensor Core instructions~\cite{nv_tensor_core}, which are more commonly used in larger neural networks because they are more efficient. 
See an example of a raw trace of the first layer in \cref{fig::trace}. For parameter extraction, \emph{preprocessing} of the traces is important as we empirically found that taking a moving average of multiple (3--5) clock cycles of the raw traces enhances the observed leakage. The effectiveness of this preprocessing can be partially explained by the uncertainty regarding warp scheduling, which does not guarantee the execution of a warp with a certain sensitive intermediate value in the same clock cycle in every trace. Therefore, the leakage might be scattered across multiple clock cycles, and the moving average helps to gather the leakages.

\subsection{Tensor Core Implementations}\label{ssec:timpl}
The implementations utilizing Tensor Cores refer to math pipelines in the GPU's SMs that accelerate GEMM operations for fixed matrix sizes. This acceleration can be achieved with different instructions for different data types and matrix sizes. 
For our 2-layer CNN, the first convolutional layer uses the \instr{IMMA.16816.S8.S8} instruction~\cite{ampere_inst}, reverse-engineered from the CUDA implementations of TensorRT. 
This instruction accelerates a matrix multiply of a matrix of size $\mathit{n\times m = 16\times 16}$ multiplied by a matrix of size $ \mathit{k\times m = 16\times 8}$ and the addition of a matrix of size $\mathit{k \times k=16\times 8}$ with signed integers \instr{S8} as weights. The first matrix contains different inputs for convolution in each row, while the second matrix is the weight matrix with columns of weights from different kernels. 
The instruction requires the cooperation of all threads in a warp, as inputs and weights are provided by the threads' registers. The different results are also written to the registers of different threads in the warp.
It is important to note that each final result written into the register file can depend on up to 16 weights, i.e., the complexity of the attack would be $16 \times 8 = 128$ bits. However, in this particular implementation, only 4 weights are loaded from each kernel for an individual \instr{IMMA.16816.S8.S8} instruction.
This would make the attack infeasible unless the inputs are chosen, i.e., by setting the appropriate inputs to zero, such that the final result depends on only one weight.

Furthermore, the implementation of the first layer is executed by 512 threads that are then divided into 16 warps when scheduled for execution. 
These warps are further divided between four SMs on the GPU, i.e., one SM gets four warps.
Notably, the four warps in each SM calculate the convolution with all the weights but different inputs. This means that theoretically, we only have to target one SM with our EM probe. However, for larger models, more SMs are likely to be required to probe.
Additionally, we still collect EM traces for all SMs to be able to compare the leakage.

\subsection{Warp-level Leakage Model}
Accurately describing the GPU's power consumption is challenging due to the many parallel threads executing. However, focusing solely on the subpartitions of the SMs, it is feasible to develop a power model that more accurately reflects the power consumption of each subpartition.
Threads in a warp physically start executing at the same time, and although the threads in a warp can diverge, highly optimized implementations try to avoid this as it incurs a performance penalty.
Therefore, it is reasonable to consider the power consumption of all the threads in a warp as they reach the writeback stage at the same time or approximately the same time. 
Our leakage model captures this idea and provides an aggregated power consumption for a warp. In a nutshell, the leakage model targets registers overwritten by an \instr{IMMA.16816.S8.S8} instruction. However, only a fraction of these registers are used in the leakage model, namely those whose value depends on the same weight. 
The leakage generated by other registers' computation is simply treated as noise. 

Let $t_x$ denote an individual sample of a trace $\mathbf{T} = [t_0, t_1, \dots, t_m]$, where $m$ denotes the number of sample in the trace and $ \mathit{0 \leq x\leq m}$.
For a warp $W_j$, if it executes an \instr{IMMA.16816.S8.S8} instruction at time sample $t_x$, the same weights are multiplied with $n$ different inputs at the same time, therefore, we build our correlation attack on the sum of the power consumption of register overwrites with new partial sums ($s_{d+1}^{i}$): 
\begin{equation}
    \mathit{P(W^x_j) = \sum_{i}^{n} HD( s^{i}_{d}, s^{i}_{d+1})},
\end{equation}
where $j= 1, \dots, k$  $\mathbf{k}$ is the number of warps and $\mathbf{n}$ is the number of parallel multiplications with the same weight but different inputs. $s_{d+1}^{i}$ denotes the partial sum result of the multiplications calculated with the $\mathit{i}$-th input row. 
%\lukasz{I modified this paragraph - have a look.}

%experiment setup, for the first layer, 
In our experimental setup, the \instr{IMMA.16816.S8.S8} instruction produces 16 different dot products for each weight vector so $n=16$. %\lukasz{these two sentences do not connect well}
Therefore, we calculate the correlation between the power predictions and the leaking sample: $\mathbf{\rho}(P(W^x_j), \mathit{t_x})$.
This is the \emph{warp-level} correlation, exploiting that the same weight is used at the same time with different inputs in a single warp. Extending this to other matrix sizes is straightforward, and the approach is weight-type agnostic as well, i.e. the same approach can be used for floating-point weights, as we show later in our LLM attack with \instr{bfloat16} weights.

\subsection{Higher-order Leakage Model}\label{ssec:holm}

The previous method only considered the leakage of intermediate values computed in parallel. However, one trace contains more information related to the same weight. This is a consequence of how neural networks work: the same weights are used in dot products with different inputs many times. For instance, in convolution, a kernel has to slide over the entire input range and produce intermediate values that depend on the same weights but different inputs.
Therefore, there is more to exploit in one trace than just a single time point for a single weight. This idea has been used before to recover input images of neural networks~\cite{poster_input}, but our approach differs in that the horizontal leakages are combined across multiple traces to recover weights, not inputs.
We can combine multiple leakages that occur in different time samples ($t_x, t_y$) related to the same weight (e.g., the results of two \instr{IMMA} instructions if they use the same weights) in a similar way that it is done for masking~\cite{chari_masking,mangard2008power}. First, we will need a \textit{combining function}~\cite{mangard2008power}; and simply the sum of the power consumptions of individual warps works well: 
\begin{equation}
    P_{comb}(W^x_{j}, W^y_{k}) = P(W^x_{j}) + P(W^y_{k}).
\end{equation}

%\footnote{In addition, the square of the sums $(P(W^x_{j}) + P(W^j_{k}))^2$ and the product $(P(W^x_{j}) \times P(W^j_{k}))$ also seem to provide similar results.}
We also need a \textit{pre-processing function}~\cite{mangard2008power} to create a new sample $t_c$ from the samples ($t_x, t_y$), and based on our experiments, the square of the sum of the samples works the best, i.e., $t_c = (t_x + t_y)^2$.

Then, we calculate the correlation of the two variables as follows: 
\begin{equation}
    {\rho}(P_{comb}(W^x_{j}, W^y_{k}), t_c).
\end{equation}
This method can be extended to more than two time samples (i.e., more than 2 \instr{IMMA} instructions), in the same way if they depend on the same weight.

%potentially 
In the results below, we recover only 8 bits (one weight at a time) by setting 3 inputs to zero and 1 random non-zero value for the eight weights in different kernels. 
For a full 32-bit attack without chosen inputs (i.e., attacking all 4 weights simultaneously), we would likely need more traces to succeed. 
An alternative could be 2 chosen-input attacks (setting only 2 inputs to zero), recovering the first two weights, then the second two, so 16 bits at a time. For the very first weight in each kernel, as also shown in previous works~\cite{DBLP:conf/iscas/YoshidaKOSF20}, the correlation might not be robust, and using the first two weights and the corresponding partial sums instead is a better strategy anyway.
Furthermore, it is important to note that partial sums in the Tensor Core units could also be attacked, but the exact calculation of these partial sums is unclear. We leave this for future work. 
Nevertheless, in this paper, for the sake of simplicity and since we merely present a proof-of-concept attack, we only target 8 bits at a time by attacking registers in the register file of subpartitions, not registers in the Tensor Cores.
%\lukasz{modified paragraph}

\subsection{Parameter Extraction Results}\label{ssec::wer}
\begin{figure}
	\centering
	\includesvg[width=\linewidth]{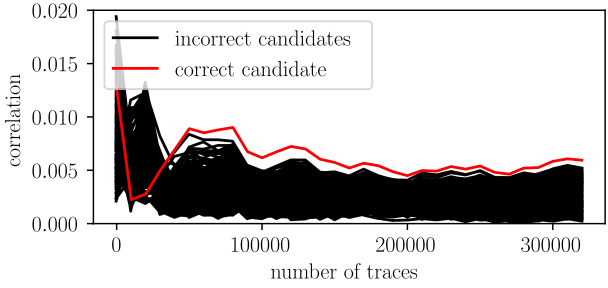}
	\caption{\label{fig::warp_level1} Warp-level correlation of eighth weight in the first kernel.}
	\end{figure}
	
\begin{figure}
\centering
\includesvg[width=\linewidth]{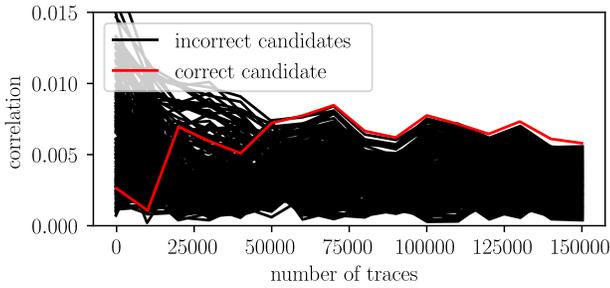}

\caption{\label{fig::warp_level2} Warp-level correlation of eighth weight in the second kernel.}
\end{figure}

\begin{figure}
\centering
\includesvg[width=\linewidth]{results/2combinedcpa_plus_leakage_plus_samples_iter.csv_corr_plot.svg}
\caption{\label{fig::2combined} Correlation with 2 warp-level intermediates combined for the eighth weight in the first kernel.}
\end{figure}
	
\begin{figure}
\centering
\includesvg[width=\linewidth]{results/3combinedcpa_plussquared_leakage_plus_samples_iter.csv_corr_plot.svg}
\caption{\label{fig::3combined} Correlation with 3 warp-level intermediates combined for the eighth weight in the first kernel.}
\end{figure}

\begin{figure}
    \centering
    \begin{subfigure}{0.9\columnwidth}
        \centering
		\includesvg[width=\linewidth]{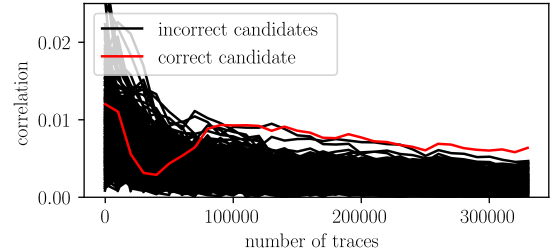}
		\caption{\label{fig::warp_level3} Warp-level correlation of the eighth weight in the third kernel.}
    \end{subfigure}%
    
    \begin{subfigure}{0.9\columnwidth}
        \centering
		\includesvg[width=\linewidth]{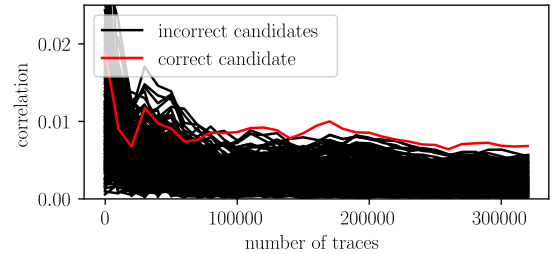}
		\caption{\label{fig::warp_level4} Warp-level correlation of the eighth weight in the third kernel.}
    \end{subfigure}
    \caption{Correlation for eighth weight in the third kernel with different warp correlations.}
\end{figure}

In this section, we provide the results for our weight extraction with different leakage models. The attacks were mounted against the first layer of the 2-layer CNN and this layer has 288 weights.
We are able to extract all the weights using our methodology.

\subsubsection{Warp-level correlation results}
In general, warp-level correlation already significantly improves the attack over BarraCUDA~\cite{horvath2025barracuda}, requiring only 300\,k traces on average to extract weights, despite the much higher clock frequency in our setup. On average, 100\,k traces are sufficient to extract weights in our attack, in contrast to millions of traces required in BarraCUDA.
For example, see \cref{fig::warp_level1} for the evolution of CPA for the eighth weight in the first kernel of the first layer. Similarly, \cref{fig::warp_level2} shows the evolution of CPA for the eighth weight in the second kernel. Lastly, \cref{fig::warp_level3,fig::warp_level4} show the results of using different warps, and hence different intermediates for the same weight.
This leakage model, which considers all 32 threads of a warp, provides a more accurate description of the power (EM) consumption, thereby reducing the number of traces. 
Furthermore, we analyzed all the active SMs on the die, and all of them leak weights, although each SM does not necessarily leak all of the weights. Nevertheless, all the weights can be extracted.

\subsubsection{Higher-order attack results}

Combining two (\cref{fig::2combined}) or three (\cref{fig::3combined}) warp-level correlations can further improve results. The leakages for these figures were collected from separate SMs, but each combined plot uses two or three intermediate values from the same SM. 
In general, a combination of intermediate values helps with faster convergence, but finding the optimal leaking spot for the EM probe is, of course, still the most important. At the same time, a particular location is restricted in the number of intermediates that leak.  
With three warp-level correlations shown in \cref{fig::3combined}, the key rank goes to 0 just after 10\,000 traces.

%\lukasz{updated, are you fine with that?}
%\lejla: Looks good to me, I think we need a version in the system without comments.
The above approach is somewhat similar to the multi-target CPA attack, which performs first-order CPA multiple times and combines the results afterwards~\cite{DBLP:conf/asiacrypt/MatherOW14}. We did not implement this approach; instead, we performed a higher-order attack. We consider investigating multi-target CPA in this context as future work.

\section{Far Field Attack}\label{sec::ffa}

\subsection{Threat Model}\label{ssec:threat}
For this attack threat model, we consider an attacker whose goal is to extract information about an LLM that is accelerated on a GPU. 
The attacker exploits the information contained in the electromagnetic radiation of the GPU.
More specifically, the attack works in two steps:
\begin{itemize}
    \item First, identify the architecture of the target LLM based on the radiated signals.
    \item Second, extract the LoRa weights of the model using the same radiated signals.
\end{itemize}
We assume that the target LLM is an open-source model fine-tuned for a specific task by a developer using LoRa~\cite{lora_weights}, and the architecture has already been identified by the attacker. Training LLMs from scratch is expensive, and for many, the only alternative is to use an open-source base model and adapt it to their specific application. 
However, if a proprietary model is targeted, then an attacker must recover all the weights.
In this work, we consider electromagnetic radiation of the GPU only in the far-field region. This assumes a weaker adversarial model than all of the existing works, where near-field physical access and sometimes semi-invasive modification (heatsink removal) of the target are required~\cite{horvath2025barracuda}.
In addition, for consumer GPUs such as the 4090, removing the heatsink for near-field access might easily cause damage without appropriate cooling equipment. In addition, even with adequate equipment, placing a near-field probe over the GPU die is much less stealthy, and the attacker still runs the risk of overheating and the GPU shutting down, potentially compromising the whole attack.

\subsection{Far-field Modeling}\label{ssec::ffmodel}

In the far field, we assume the clock signal is a carrier amplitude-modulated with weight-related information. Formally, we can simply express the signal of interest as

\begin{equation}
    s(t)=A(t)\cos(2\pi ft + \phi),
\end{equation}where $t$, $f$, and $\phi$ denote time, frequency, and phase, respectively. $A(t)$ represents the data that we are interested in.
For the 4090 RTX GPU in our experiments, $f=$ 2\,565\,MHz. An adversary has two options to measure this RF signal: 
\begin{enumerate}
    \item Direct sampling: Same as in the near-field experiments, i.e., use an oscilloscope to measure the whole spectrum up to a certain frequency. For very high frequencies, this method requires very expensive equipment with enough bandwidth and sampling rates. With high end chips like consumer GPUs, the frequency can be well over 2\,GHz which requires very expensive measurement equipment. 
    %with a bandwidth of at least 4\,Ghz.
    %Lukasz: the above is not necessarily true
    \item Zero-IF sampling: This method allows the attacker to shift and downconvert the high-frequency signal to baseband, essentially keeping only $A(t)$. 
    Therefore, an ADC with a very high sampling rate is not required, and SDRs are an appropriate choice for this application. Naturally, there might be more frequencies that carry relevant information, and this method is restricted to a single frequency with a narrow bandwidth. However, our experiments prove that the clock signal carries weight-dependent information.
\end{enumerate}

\subsection{Experiment Setup}\label{ssec:exp}

\begin{figure}[htbp]
\includegraphics[width=0.48\textwidth]{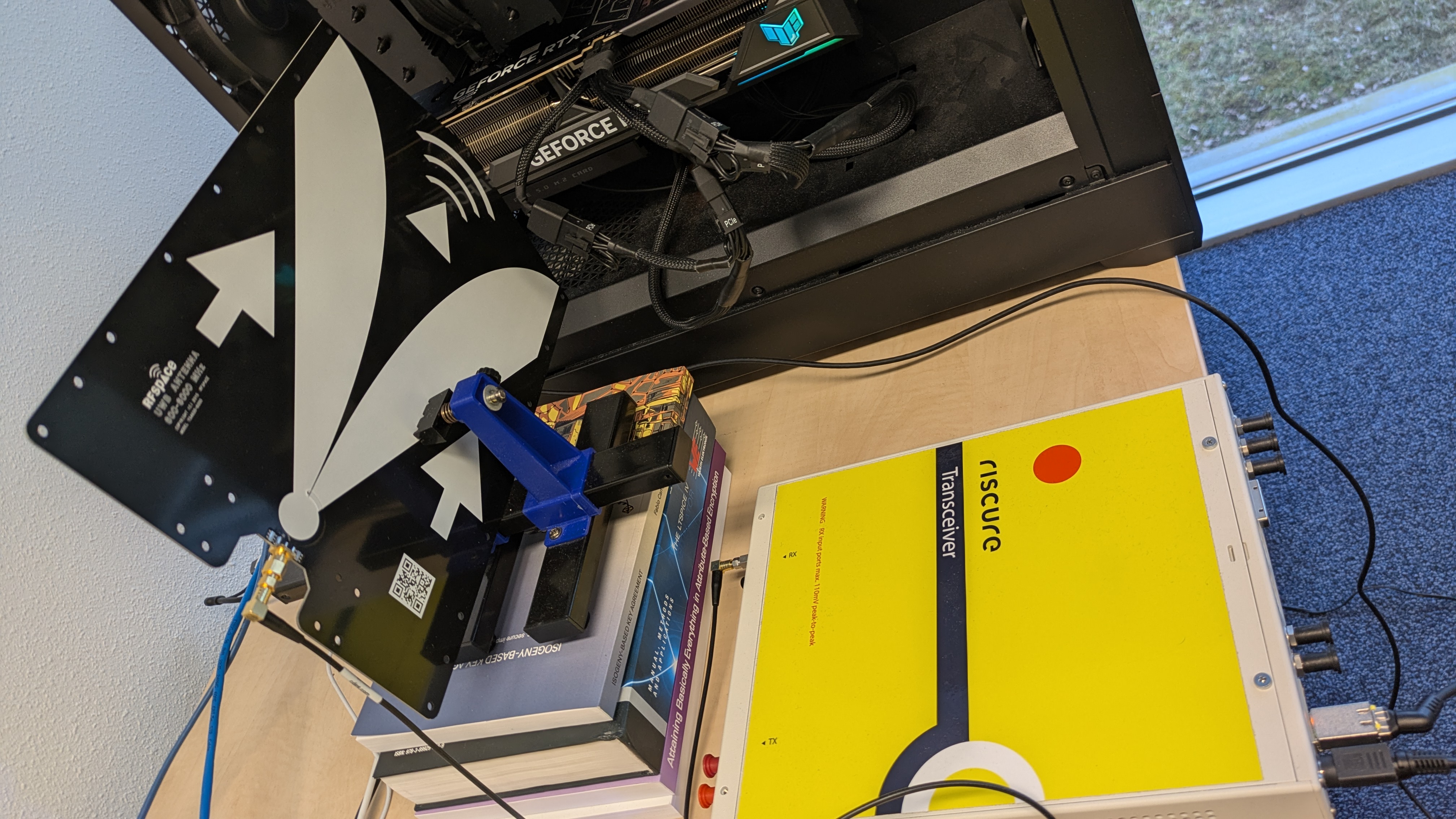}
\caption{Measurement setup 25 cm away from the GPU die.}
\label{fig::setup1}
\end{figure}

\begin{figure}[htbp]
\includegraphics[width=0.48\textwidth]{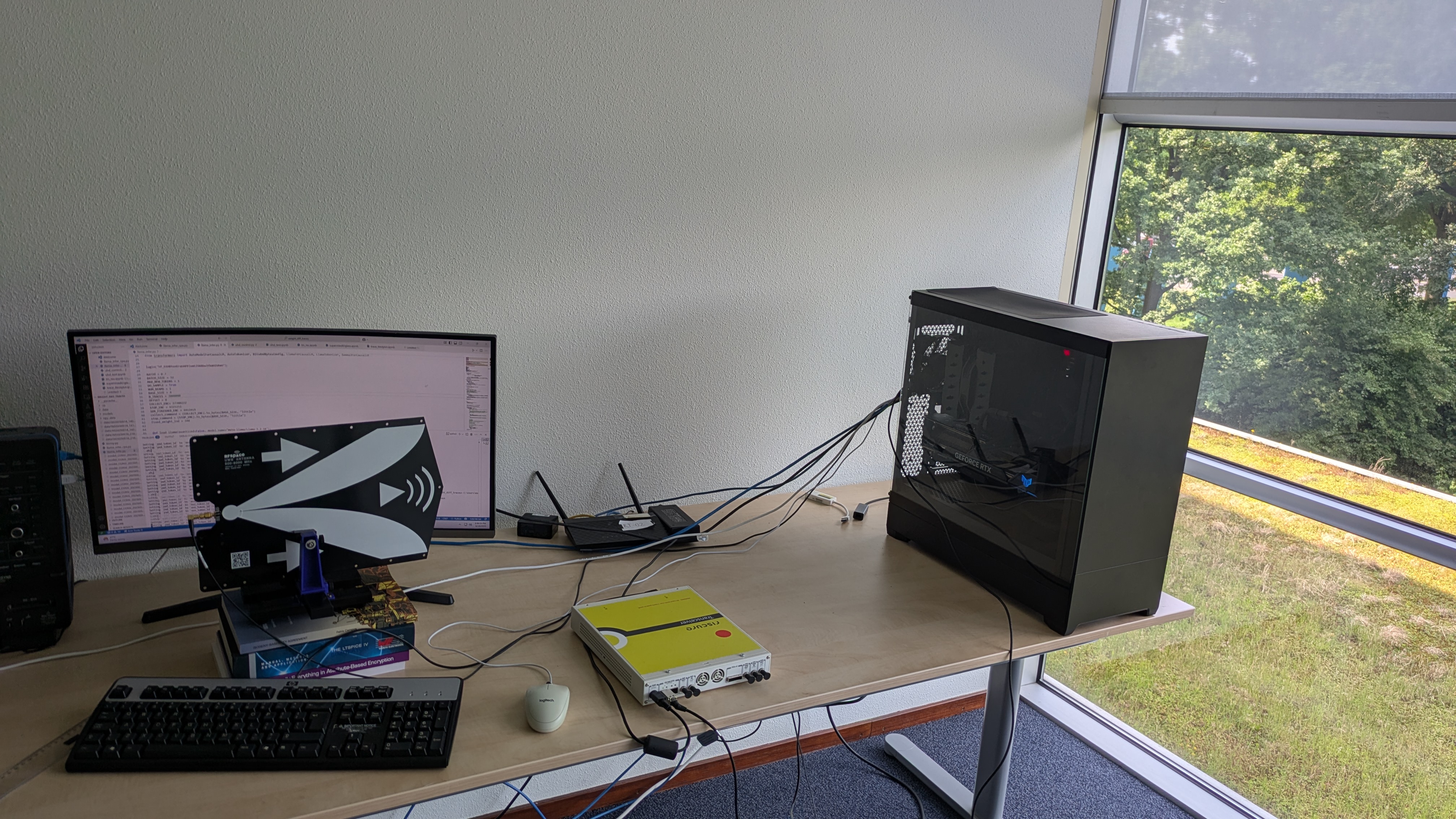}
\caption{Measurement setup 100 cm away from the GPU die with glass in between. Since glass has a higher dielectric constant than air, there is an expected reduction in leakage.}
\label{fig::setup2}
\end{figure}

In order to measure electromagnetic radiation in the far-field, we use the RFSpace Vivaldi antenna~\cite{rfspace}
and Ettus X310 Software Defined Radio (SDR)~\cite{ettus_sdr} as shown in \cref{fig::setup1}, similarly to ~\cite{farfield_clairvoyance}.

The center frequency of the SDR is set to the GPU's core clock boost frequency that is $f_c=$ 2\,565\,MHz without the OC mode for the Asus TUF Gaming NVIDIA GeForce RTX 4090 OC Edition~\cite{tuf_4090}. However, the GPU switches frequency based on load, so higher frequencies are also possible. Still, as discussed later, the center frequency remains at 2565\,MHz in all experiments. The SDR's bandwidth and sampling rate are 2\,MHz and 4\,MHz, respectively. This attack is non-invasive, as no modifications are made to the target PCB. We provide more details on various aspects of the setup below.

\subsubsection{Antenna Positioning}
In our case, the boundary between the near and far fields with $\mathit{R = \frac{2 \times D^{2}}{\lambda}}$, where $\mathit{\lambda = \frac{1}{f_c} \approx 12 cm}$ and $D= 2,5 cm$ is the largest dimension of the radiating source, the GPU die. This gives us $\mathit{R = \frac{2 \times 0,025^{2}}{0.12}} \approx 0.01042 m$. Therefore, according to this formula, the far-field region starts approximately 1 cm away from the GPU die. However, it is not necessarily the die that radiates, so this measurement might not be accurate in establishing the far-field boundary. Nevertheless, our goal is to show that unfettered near-field access may not be required to observe weight-related leakage in the EM signals.

We experimented with different distances, such as 25 and 100 \,cm for the antenna distance from the GPU.  
The antenna in \cref{fig::setup1} is positioned 25\,cm from the center of the die of the 4090 GPU with the glass casing removed, while \cref{fig::setup2} shows the antenna 100\,cm away from the GPU die with glass as an obstacle.

\subsubsection{Center Frequency}
The center frequency of the SDR is one of the most important parameters to consider for the success of the attack. According to our tests, the radiation at the frequency of the GPU's DRAM is much stronger than the radiation at the GPU's core clock frequency, and this also aligns with the findings in previous work~\cite{liang2022clairvoyance}. However, our goal is weight extraction, i.e., finding fine-grained data dependencies in the radiated signals, not just operation-dependent signals. This may be more likely with the GPU core clock signal modulated with data, allowing us to extract weights due to the closer proximity of the clock signal and the data. On the other hand, a specialized antenna is likely required to successfully mount the attack from greater distances, as these signals are much weaker.   
%\lukasz{maybe it is good to list already what is used here?}

\subsection{GPU Llama 3.2 1B Implementation}\label{ssec::llama_impl}
We use the Huggingface (4.51.3) Python library to run the Llama 3.2 1B model~\citep{grattafiori2024llama3herdmodels} on the GPU. The model weights are \instr{bfloat16}.
The GPU has specific math pipelines to accelerate matrix multiplications for different data types, including \instr{bfloat16}. The GPU implementation of the first layer uses the \instr{HMMA.1688.F32.BF16} instruction to implement matrix multiplications of the embeddings and the $\mathbf{W_q}, \mathbf{W}_k, \mathbf{W_v}$ weight matrices.
As the \instr{.F32.BF16} suffix suggests, the dot products are calculated by multiplying \instr{bfloat16} numbers and accumulating them in \instr{float32}.  This information is crucial for CPA-based side-channel attacks to accurately model leakage.

\subsection{CPA Attack Complexity}\label{ssec::cpa_comp}
Depending on the implementation and the targeted intermediate results, the complexity of the attack can vary. The architecturally visible registers of the GPU are overwritten by the results of the Tensor Core instructions, such as \instr{HMMA}. As these instructions can accelerate multiplications with different matrix sizes, the following are influenced: 
\begin{enumerate}
    \item the number of registers that are overwritten,
    \item the number of weights used in the final results that are used to overwrite the registers.
\end{enumerate}

For a register in the register file, the result of a Tensor Core instruction can depend on multiple weights, based on the available instruction formats~\cite{warp_matrix}. For Llama 3.2 1B and its implementation on the RTX 4090 with the \instr{HMMA.1688.F32.BF16} instruction, the number of weights used in the dot product is 8, so the result in a register depends on 8 weights. 

In other words, the stored sum $s_{reg}$ in the register is the dot product between the weights $w_i$ and embeddings $x_i$:

\begin{equation}
s_{reg} = \sum_{i=1}^8 w_i*x_i.
\end{equation}
Since $s_{reg}$ depends on 8 weights, a naive attack would be predicting all the 8 16-bit weights at the same time, meaning a complexity of $8 \times 16 = 128 $ bits.
However, the sum can be rewritten as
\begin{equation}
s_{reg} = w_8*x_8 + \sum_{i=1}^7 w_i*x_i = s_1 + s_2,
\end{equation}
where $s_1$ is \instr{float32} and $s_2$ is also \instr{float32}, but $s_1$ depends on only 1 weight. If $s_2$ is fixed, but not necessarily zero, then an attacker can build a leakage model that predicts $s_1$ and $s_2$ at the same time. This means that an attacker has to predict the fixed $s_2$ (32 bits) while also predicting $s_1$, which depends on one 16-bit weight only. This idea is similar to the attacks presented on the MixColumns operation for AES implementations~\cite{aes_chosen_input, cryptoeprint:2019/343}. Fixing the state for $s_2$, the attacker can recover the weight that $s_1$ depends on without predicting the other weights. Furthermore, depending on the number of weights in the dot product, the $s_2$ state will be bounded, so the complexity will be at most 32 bits. 

In general, to attack $w_k$ in a dot product with $n$ weights: 
\begin{equation}
    s_{reg} = w_k*x_k + \sum_{i=1, i\neq k}^n w_i*x_i = s_1 + s_2.
\end{equation}
In the sum, $s_1$ is just the multiplication result for the weight $w_k$ as before, and $s_2$ is the sum of all other multiplications in the dot product. This attack can be parallelized for each weight, or can be mounted sequentially to incorporate already extracted weights. The sequential approach is as follows. If $m-1$ weights are already extracted, and the targeted weight is $w_m$, 
\begin{equation}
s_{reg} = w_m*x_m + \sum_{i=m+1}^n w_i*x_i + \sum_{i=0}^{m-1} w_i*x_i = \\ s_1 + s_2 + C,
\end{equation}
where $C$ is the sum of the already extracted weights multiplied by the corresponding inputs. 
%\lukasz{this paragraph is not clear to me} \peter{added a paragraph}\lukasz{now clear!}

In addition, due to the distribution of weights in neural networks, the full range of \instr{bfloat16} does not have to be considered: for every weight $w$ the following holds: $1\mathrm{e}{-10} < |w| <1$ in the first layer of Llama 3.2 1B. The number of \instr{bfloat16} weights in this range is 5\,106, so only $\log_25\,106 \approx 12.318$ bits.

Therefore, the attack complexity to recover a single 16-bit \instr{bfloat16} weight is at most $44$ bits, but this has to be repeated for every weight. Since LoRa only requires extracting a small subset of weights, the 44-bit complexity does not make the attack impractical for adversaries with adequate computing resources. 

Note that this extends to all cases where we have a result that leaks and depends on multiple weights, regardless of data type, i.e., the same applies to integer parameters. Therefore, the attack does not require chosen-zero inputs; it only requires chosen-fixed inputs for the 32-bit state.  For instance, if only the final result of a dot product between an embedding vector and a a column of a projection matrix (e.g., $\mathbf{W_k}$) leaks, this attack still enables an adversary to recover all the weights used in the dot product.
However, for LLMs, individual embedding vectors are fixed, so highly advanced techniques (e.g., laser fault attacks) might be required to exploit near field access for this to work. The idea would be to force all the bits of $s_1$ to a constant value for every trace. However, this may not be very practical and the attacker would have to attack the individual multiplications in the Tensor Cores directly. 

\subsection{Input-related Leakage}\label{ssec::ileakage}
\begin{figure}[h!]
\centering
\includegraphics[width=0.9\linewidth]{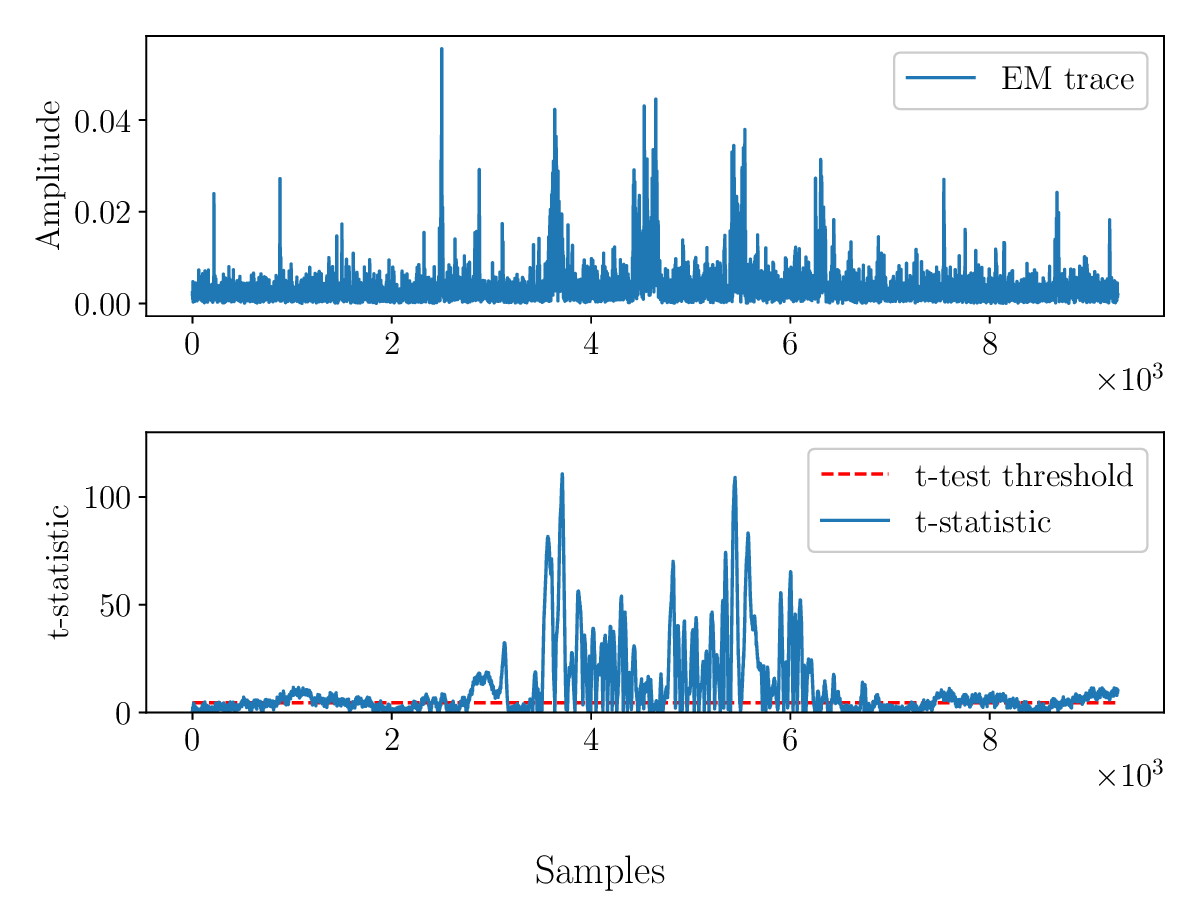}
\caption{Fixed vs. random input TVLA plot in the first layer of Llama 3.2 1B, with batch size 32, for the first generated token, with 50\,000 traces. The same input is repeated 32 times in the batch.}
\label{fig::llama_ttest_input}
\end{figure}

We also investigated input-related leakage using fixed vs. random TVLA as shown in \cref{fig::llama_ttest_input}. In this test, we set every input encoding to random vs. fixed, hence the large TVLA peaks throughout the trace. In general, this shows that even in the far field, there is leakage associated with the inputs. 
%\lukasz{why is it important? I think that it should be clarified.}

\subsection{Weight-related Leakage}\label{ssec::wleakage}
\begin{figure}[htbp]
\centering
\includegraphics[width=0.9\linewidth]{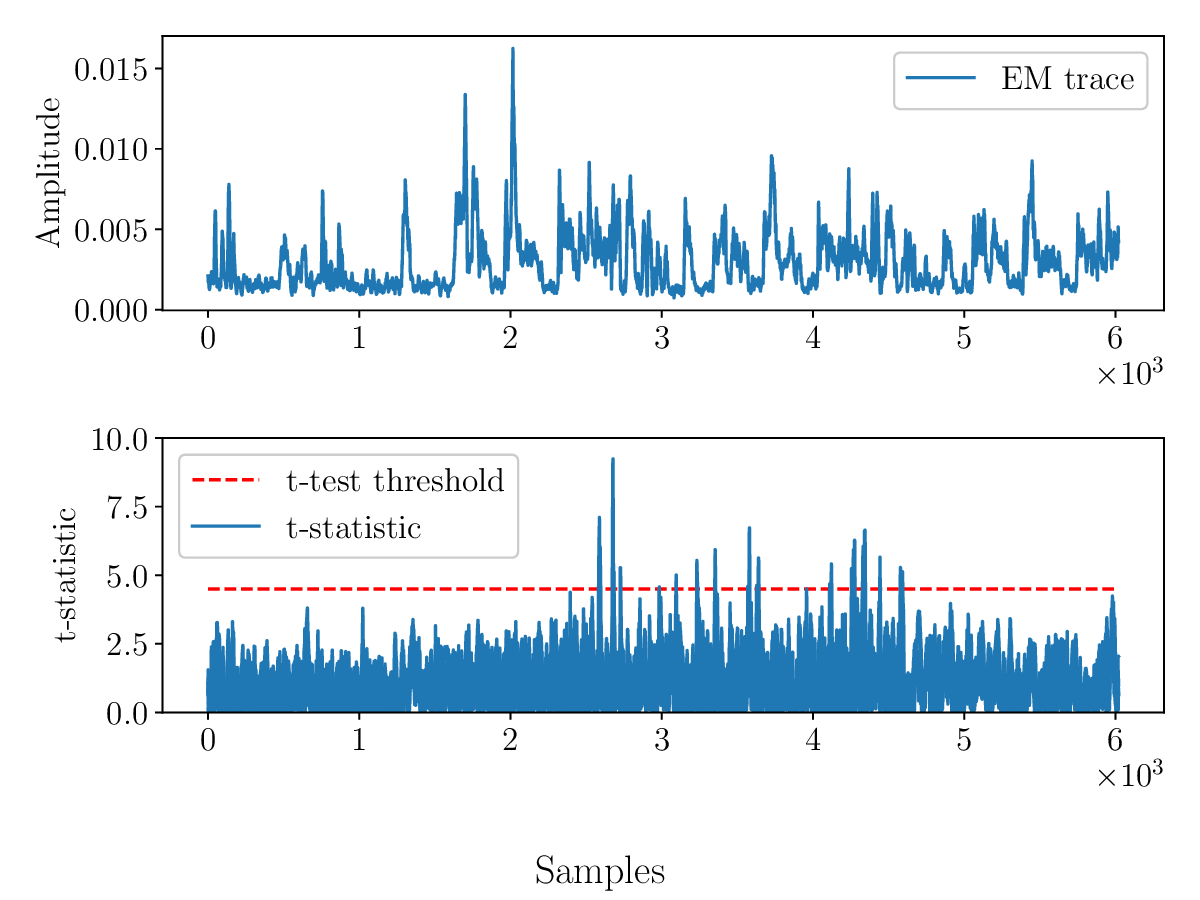}
\caption{Fixed vs. random weight TVLA plot in the first two layers of Llama 3.2 1B, with batch size 32, for the first generated token, with 300\,000 traces. The first layer's first weight in $\mathbf{W_q}$ is the randomized weight. The t-test plot is an aggregated result of aligning the traces at different time samples due to the jitter present in the traces.}
\label{fig::llama_ttest_b32}
\end{figure}

\begin{figure}[htbp]
\centering
\includegraphics[width=0.9\linewidth, height=6cm]{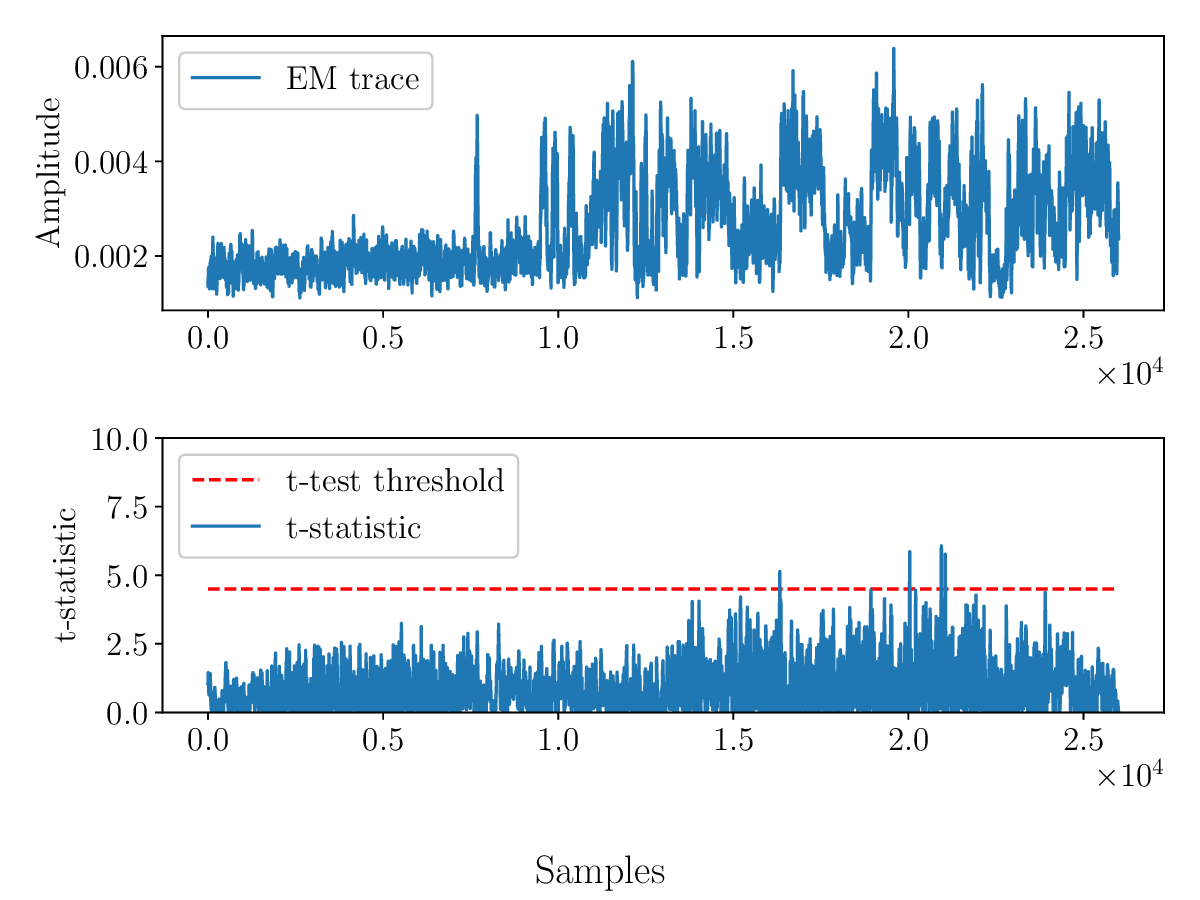}
\caption{Fixed vs. random weight TVLA plot in the first layer of Llama 3.2 1B, with batch size 128, for the first generated token, with 300\,000 traces. The first layer's first weight in $\mathbf{W_k}$ is the randomized weight. The $t$-test plot is an aggregated result of aligning the traces at different time samples due to the jitter present in the traces. The batch size is 128, with the same input repeated 128 times. The antenna is 25 cm away, with no glass between.}
\label{fig::llama_ttest_b128}
\end{figure}

In order to test the feasibility of weight extraction, we use random vs. fixed TVLA~\cite{schneider2015leakage} to assess whether there is weight-related leakage present in the far-field traces. To that end, we run multiple experiments to test the leakage of weights in the matrices $\mathbf{W}_q, \mathbf{W}_k, \mathbf{W}_v$. In addition, we run the experiments with different batch sizes. Results are shown in \cref{fig::llama_ttest_b32} and \cref{fig::llama_ttest_b128}. 
%\lukasz{It should be clarified how these results are chosen (i.e., "some"). }
For a batch size of 32, the GPU's clock frequency remains at the default boost frequency during the execution of the forward pass of the Llama model. For a batch size of 128, the clock frequency jumps to 2\,865 MHz during the forward pass execution. %\todo{Is that just voltage scaling? what if you hot start?}
Despite the differences in clock frequencies, \cref{fig::llama_ttest_b32} and \cref{fig::llama_ttest_b128} show leakage captured at a center frequency of 2\,565 MHz. This illustrates that even though the GPU clock frequency is different in these scenarios, the GPU Boost Clock frequency (2\,565Mhz) still leaks even for higher batch sizes executed on a higher clock frequency due to increased GPU load.
However, the amplitude of the radiated signal for batch size 128 is slightly lower than that of batch size 32 when measured at the 2\,565 MHz frequency. This behavior suggests that clock frequency jumps between these intervals.
When measuring at the indicated 2\,865 MHz frequency according to GPU monitoring tools, the radiated signal at 2\,865 MHz during the forward pass is negligible. 
\subsection{Number of Tokens and Batch Size}\label{ssec::ltokenbatch}
\begin{figure}[h!]
\centering
\includegraphics[width=0.9\linewidth]{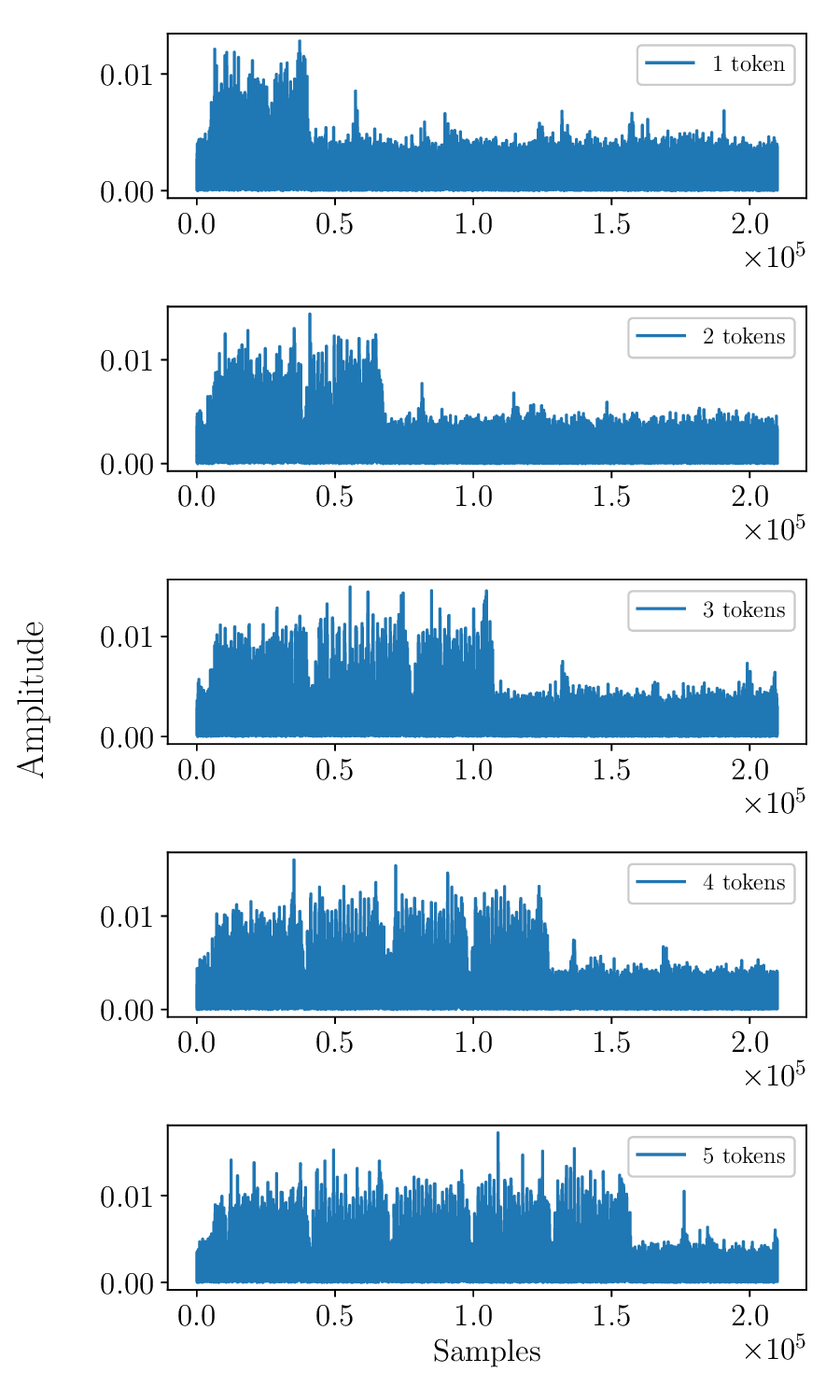}
\caption{The number of generated tokens is clearly visible by capturing the emitted radiation. The batch size is set to 32. The number of generated tokens is 1 (top) to 5 (bottom).}
\label{fig::llama_n_tokens}
\end{figure}

One generated token corresponds to one forward pass through the model, and this is also clearly visible in the EM traces shown in \cref{fig::llama_n_tokens}. It is also visible that the first token generation always takes more time than the subsequent tokens; this could partially be explained by caching.

\begin{figure}[htbp]
\centering
\includegraphics[width=0.9\linewidth]{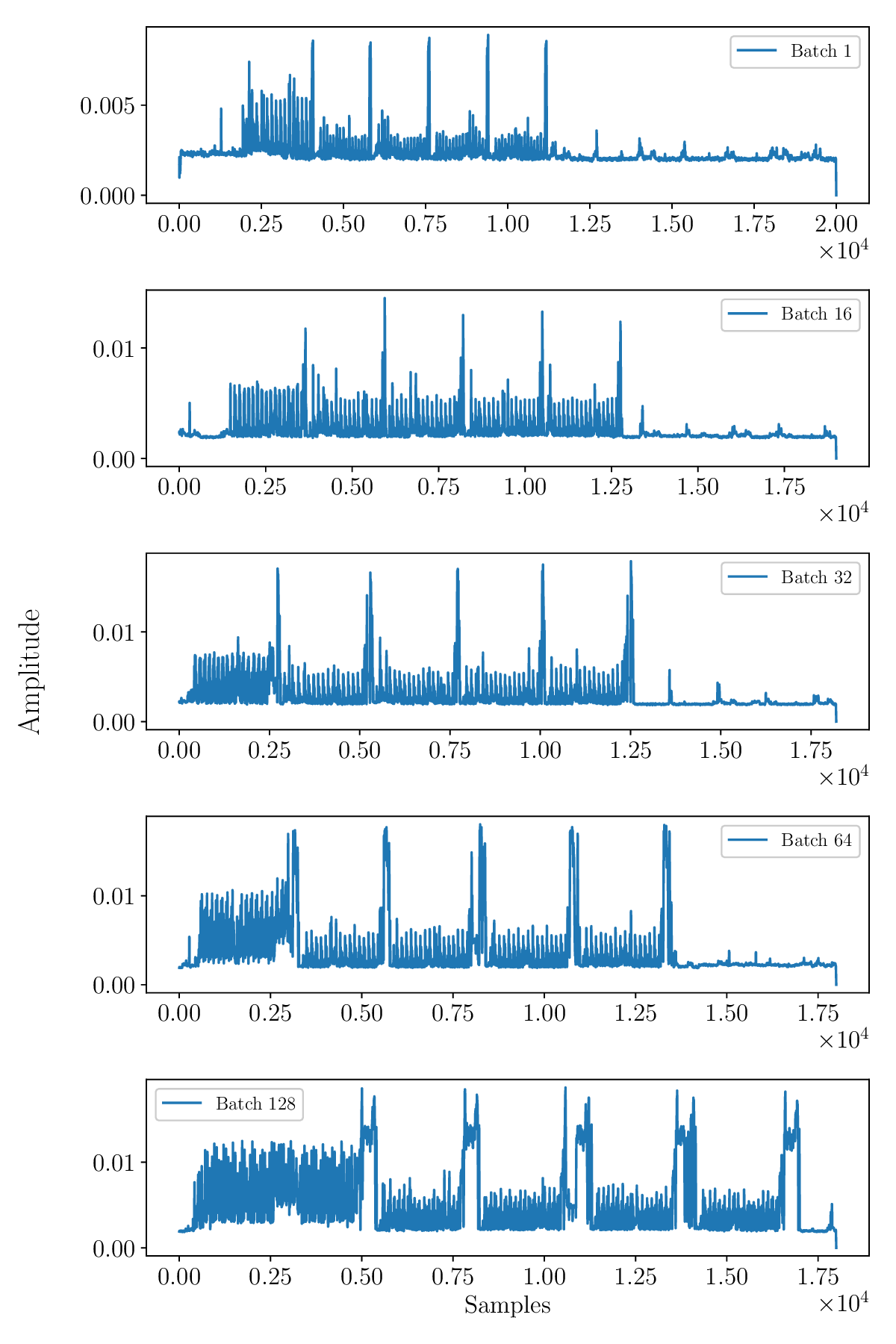}
\caption{Impact of batch size on latency, with the number of generated tokens set to five. The number of generated tokens is clearly visible, while a higher batch size provides increased latency, albeit not linearly.}
\label{fig::llama_batch}
\end{figure}

The impact of batch size is shown in \cref{fig::llama_batch}, where increased latency is clearly correlated with increased batch size. Although the latency increase is visible due to increased batch size, it is not a linear relationship, and this behaviour depends on the implementation and the GPU's resources.

\subsection{Parameter Extraction Results}\label{ssec::ff_cpa}
As a proof-of-concept, we demonstrate the extraction parameters in far-field, assuming we know the first 7 weights in the matrices $\mathbf{W_q}, \mathbf{W_k}, \mathbf{W_v}$, and we target the 8th weight, so we only have 5\,106 candidates for CPA.
By building the leakage models on the power consumption of individual warps, we can reach low key ranks for weights from $\mathbf{W}_q, \mathbf{W}_k, \mathbf{W}_v$ from 100 cm away from the GPU die through glass. We average 500 traces for each input, and we use 4\,000 different inputs, i.e., the number of collected traces is 2 million.

The results are shown in \cref{sec::app} for the 8th weight in each of the matrices $\mathbf{W_q}, \mathbf{W_k}, \mathbf{W_v}$. In general, summing up the power consumption related to the same weight in the warp matrix-multiply instruction is a viable strategy. At 25 cm away from the die without glass in between, collecting only 1M traces with averaging already provides better results than 2M traces collected 100 cm away from the die with glass in between. 
Overall, we still expect better results with more traces, but our results already demonstrate that weight leakage can be observed in far field as well.

\section{Discussion}\label{sec::disc}
\subsection{Limitations}
Although we showed that our approach across different GPU architectures and different neural network types (CNN, LLM) can work, the results in the far field are limited in practice.
Furthermore, full model extraction, i.e., extracting the weights of all layers, remains a challenging problem largely due to the computational resources required to compute every weight. In addition, because LLMs use an embedding layer, chosen-input attacks will not work against LLM-based architectures. Therefore, an adversary will have to target each multiplication in the Tensor Cores for LLMs. 
In addition, the far-field attack is a proof-of-concept demonstrating that leakage can be observed in the far field as well, although exploiting it is not yet practical. However, the observation regarding LoRA weights also stands for near-field attacks, for which we have derived a much more efficient attack than previous work~\cite{horvath2025barracuda}.

\subsection{Quantization Leakage}

Because quantization schemes are many-to-one functions, an adversary can already deduce important information about the weights before launching a side-channel attack if it is known that the weights were quantized. 
When a set of weights of a trained model is converted from FLOAT32 to a signed integer with precision $b$, affine quantization is used~\cite{DBLP:journals/corr/abs-2004-09602} where a floating point weight tensor's ${W_{\mathit{FLOAT32}}}$ elements are multiplied with a quantization scale $q_s$. The scale depends on the range of quantization, where it is common to use the absolute maximum value of the weight tensor to derive the scale:
$q_s = \frac{2^b-1}{\max{W_{\mathit{FLOAT32}}^{abs}}}$. 
However, due to how the scale is derived, one of the quantized values will necessarily map to either $\max{W_{\mathit{FLOAT32}}}$ or $\min{W_{\mathit{FLOAT32}}}$. Therefore, an adversary already has information about one of the weights. 
Furthermore, it is common to use blockwise quantization, which splits a weight tensor ${W_{\mathit{FLOAT32}}}$ into fixed blocks and quantizes them separately. 
Therefore, if the quantization block size is also known to the attacker, as can be the case with open-source models, the attacker already has knowledge of each block's weights.
%a quantization scale $q_s = %\frac{max(W_f)-min(W_f)}{}$ is used for the mapping~.  

\subsection{Mitigation}
For near-field attacks, the countermeasures mentioned in previous work~\cite{batina2019csi, horvath2025barracuda}, such as masking or shuffling, could significantly hinder an adversary's ability to extract weights. The containment of EM radiation by shielding in the near field is more difficult to achieve, as in this scenario it is assumed that an adversary has unfettered physical access to the target. Nevertheless, (metallic) shielding would solve the leakage problem in the far-field without the need for countermeasures such as masking or shuffling.

\section{Conclusion}\label{sec::conc}
In this paper, we analyzed the electromagnetic side-channel leakage of GPUs with Tensor Core implementations by capturing EM radiation in the near and far fields. We targeted the results of the Tensor Core instructions in the register files of the Streaming Multiprocessors and provided more accurate leakage models for the energy consumption of individual warps by introducing the \emph{warp-level} model, resulting in attacks that are more efficient than the state of the art in the near field. To further improve our near-field attack, we introduced \emph{higher-order} attacks for neural networks to exploit the fact that multiple intermediate values, processed at different times, depend on the same weights. 

In the far field, we demonstrated weight leakage by mounting a proof-of-concept attack on an LLM, although its practical use is currently limited. 
We performed experiments through a glass obstacle, which reduced leakage, but there is still an observable correlation. We also analyzed the complexity of CPA in the presence of fixed inputs that can make the attack feasible in practice. Nevertheless, DNN implementations, even for CNNs, remain challenging to attack because they require applying chosen inputs to make the attacks feasible.

\section*{Acknowledgements}\label{sec::ack}
This research was supported by:
the Dutch Research Council (NWO) through the PROACT project (NWA.1215.18.014);
the TTW PREDATOR project 19782;
the CiCS project of the research programme Gravitation under the grant 024.006.037;
an ARC Discovery Project number DP210102670; %Yuval
the Deutsche Forschungsgemeinschaft (DFG, German Research Foundation) under Germany's Excellence Strategy - EXC 2092 CASA - 390781972 and through project number 560392681.  %Yuval
%the MV Impact Ai-SecTools (VJ02010010). 
This research was also partly supported by the Horizon Europe Project Cybersecurity Certification and Assessment Tools (CCAT, 101225878) project.
%Lukasz 

%\appendix
%\input{tex/appendix}



\bibliographystyle{IEEEtranSN}
\bibliography{bibliography}

@inproceedings{batina2019csi,
  title={{CSI}--{NN}: Reverse Engineering of Neural Network Architectures Through Electromagnetic Side Channel},
  author={Batina, Lejla and Bhasin, Shivam and Jap, Dirmanto and Picek, Stjepan},
  booktitle={USENIX Security},
  pages={515--532},
  year={2019}
}

@inproceedings{horvath2025barracuda,
  title={BarraCUDA: Edge GPUs do Leak DNN Weights},
  author={Horvath, Peter and Chmielewski, Lukasz and Weissbart, Leo and Batina, Lejla and Yarom, Yuval},
  booktitle={USENIX Security},
  year={2025}
}

@inproceedings{horvath2024sok,
  title={{S}o{K}: {N}eural {N}etwork {E}xtraction through {P}hysical {S}ide {C}hannels},
  author={Horvath, Peter and Lauret, Dirk and Liu, Zhouran and Batina, Lejla},
  booktitle={USENIX Security},
  year={2025}
}

@INPROCEEDINGS{farfield_clairvoyance,

  author={Liang, Sisheng and Zhan, Zihao and Yao, Fan and Cheng, Long and Zhang, Zhenkai},

  booktitle={2022 IEEE Security and Privacy Workshops (SPW)}, 

  title={Clairvoyance: Exploiting Far-field EM Emanations of GPU to "See" Your DNN Models through Obstacles at a Distance}, 

  year={2022},

  volume={},

  number={},

  pages={312-322},

  doi={10.1109/SPW54247.2022.9833894}}

@article{lora_weights,
  author       = {Edward J. Hu and
                  Yelong Shen and
                  Phillip Wallis and
                  Zeyuan Allen{-}Zhu and
                  Yuanzhi Li and
                  Shean Wang and
                  Weizhu Chen},
  title        = {LoRA: Low-Rank Adaptation of Large Language Models},
  journal      = {CoRR},
  volume       = {abs/2106.09685},
  year         = {2021},
  url          = {https://arxiv.org/abs/2106.09685},
  eprinttype    = {arXiv},
  eprint       = {2106.09685},
  bibsource    = {dblp computer science bibliography, https://dblp.org}
}

@inproceedings{NIPS2017_3f5ee243,
 author = {Vaswani, Ashish and Shazeer, Noam and Parmar, Niki and Uszkoreit, Jakob and Jones, Llion and Gomez, Aidan N and Kaiser, \L ukasz and Polosukhin, Illia},
 booktitle = {Advances in Neural Information Processing Systems},
 editor = {I. Guyon and U. Von Luxburg and S. Bengio and H. Wallach and R. Fergus and S. Vishwanathan and R. Garnett},
 pages = {},
 publisher = {Curran Associates, Inc.},
 title = {Attention is All you Need},
 url = {https://proceedings.neurips.cc/paper_files/paper/2017/file/3f5ee243547dee91fbd053c1c4a845aa-Paper.pdf},
 volume = {30},
 year = {2017}
}

@misc{amd_hw,
  title = {{AMD GPU} hardware features},
  howpublished = {\url{https://rocm.docs.amd.com/projects/HIP/en/latest/reference/hardware_features.html}},
  note = {Accessed: 2024-10-30}
}

@misc{rfspace,
  title = {{RFSPACE TSA600} {A}ntenna},
  howpublished = {\url{https://rfspace.com/RFSPACE/Antennas_files/TSA600.pdf}},
  note = {Accessed: 2025-02-12}
}

@misc{ettus_sdr,
  title = {{Ettus Research X}310 {SDR}},
  howpublished = {\url{https://www.ettus.com/all-products/x310-kit/}},
  note = {Accessed: 2025-01-10}
}

@misc{tuf_4090,
  title = {{A}sus {TUF} {G}aming {NVIDIA} {G}e{F}orce {RTX} 4090 {OC} {E}dition},
  howpublished = {\url{https://www.asus.com/motherboards-components/graphics-cards/tuf-gaming/tuf-rtx4090-o24g-gaming/}},
  note = {Accessed: 2025-04-30}
}

@inproceedings{ffaes,
author = {Wang, Ruize and Wang, Huanyu and Dubrova, Elena},
title = {Far Field EM Side-Channel Attack on AES Using Deep Learning},
year = {2020},
isbn = {9781450380904},
publisher = {Association for Computing Machinery},
address = {New York, NY, USA},
url = {https://doi.org/10.1145/3411504.3421214},
doi = {10.1145/3411504.3421214},
booktitle = {Proceedings of the 4th ACM Workshop on Attacks and Solutions in Hardware Security},
pages = {35–44},
numpages = {10},
location = {Virtual Event, USA},
series = {ASHES'20}
}

@inproceedings{ffaes2,
author = {Wang, Ruize and Wang, Huanyu and Dubrova, Elena and Brisfors, Martin},
title = {Advanced Far Field EM Side-Channel Attack on AES},
year = {2021},
isbn = {9781450384025},
publisher = {Association for Computing Machinery},
address = {New York, NY, USA},
url = {https://doi.org/10.1145/3457339.3457982},
doi = {10.1145/3457339.3457982},
booktitle = {Proceedings of the 7th ACM on Cyber-Physical System Security Workshop},
pages = {29–39},
numpages = {11},
location = {Virtual Event, Hong Kong},
series = {CPSS '21}
}

@ARTICLE{near-far-field,

  author={Johnson, R.C. and Ecker, H.A. and Hollis, J.S.},

  journal={Proceedings of the IEEE}, 

  title={Determination of far-field antenna patterns from near-field measurements}, 

  year={1973},

  volume={61},

  number={12},

  pages={1668-1694},

  doi={10.1109/PROC.1973.9358}}

@article{liu2020screen,
  title={Screen Gleaning: A Screen Reading {TEMPEST} Attack on Mobile Devices Exploiting an Electromagnetic Side Channel},
  author={Liu, Zhuoran and Samwel, Niels and Weissbart, L{\'e}o and Zhao, Zhengyu and Lauret, Dirk and Batina, Lejla and Larson, Martha},
  journal={NDSS},
  year={2021}
}

@article{daemen1999aes,
  title={{AES} proposal: Rijndael},
  author={Daemen, Joan and Rijmen, Vincent},
  year={1999},
  publisher={Gaithersburg, MD, USA}
}

@inproceedings{liang2022clairvoyance,
  title={Clairvoyance: {E}xploiting Far-field {EM} Emanations of {GPU} to" See" Your DNN Models through Obstacles at a Distance},
  author={Liang, Sisheng and Zhan, Zihao and Yao, Fan and Cheng, Long and Zhang, Zhenkai},
  booktitle={2022 IEEE Security and Privacy Workshops (SPW)},
  pages={312--322},
  year={2022},
  organization={IEEE}
}

@inproceedings{gilbert2011testing,
  title={A testing methodology for side-channel resistance validation},
  author={Gilbert Goodwill, Benjamin Jun and Jaffe, Josh and Rohatgi, Pankaj and others},
  booktitle={NIST non-invasive attack testing workshop},
  volume={7},
  pages={115--136},
  year={2011}
}

@inproceedings{elibol2012realistic,
  title={Realistic eavesdropping attacks on computer displays with low-cost and mobile receiver system},
  author={Elibol, F{\"u}rkan and Sarac, U{\u{g}}ur and Erer, I{\c{s}}in},
  booktitle={EUSIPCO},
  pages={1767--1771},
  year={2012},
}

@article{hongxin2009recognition,
  title={Recognition of electro-magnetic leakage information from computer radiation with {SVM}},
  author={Hongxin, Zhang and Yuewang, Huang and Jianxin, Wang and Yinghua, Lu and Jinling, Zhang},
  journal={Computers \& Security},
  volume={28},
  number={1-2},
  pages={72--76},
  year={2009},
  publisher={Elsevier}
}

@misc{tensorrt,
  title = {NVIDIA {T}ensor{RT}},
  howpublished = {\url{https://developer.nvidia.com/tensorrt}},
  note = {Accessed: 2025-04-25}
}

@misc{ampere_inst,
  title = {Ampere instruction set},
  howpublished = {\url{https://docs.nvidia.com/cuda/cuda-binary-utilities/index.html#nvidia-ampere-gpu-and-ada-instruction-set}},
  note = {Accessed: 2025-02-07}
}

@misc{warp_matrix,
  title = {Warp matrix functions},
  howpublished = {\url{https://docs.nvidia.com/cuda/cuda-c-programming-guide/#warp-matrix-functions}},
  note = {Accessed: 2025-05-07}
}

@misc{cryptoeprint:2019/343,
      author = {Aurelien Vasselle and Antoine Wurcker},
      title = {Optimizations of Side-Channel Attack on {AES} {MixColumns} Using Chosen Input},
      howpublished = {Cryptology {ePrint} Archive, Paper 2019/343},
      year = {2019},
      url = {https://eprint.iacr.org/2019/343}
}

@article{DBLP:journals/corr/abs-2004-09602,
  author       = {Hao Wu and
                  Patrick Judd and
                  Xiaojie Zhang and
                  Mikhail Isaev and
                  Paulius Micikevicius},
  title        = {Integer Quantization for Deep Learning Inference: Principles and Empirical
                  Evaluation},
  journal      = {arXiv preprint arxiv:2004.09602},
}

@misc{orin_soc,
  key={osoc},
  title = {Jetson Orin Nano Devkit},
  howpublished = {\url{https://developer.nvidia.com/embedded/learn/get-started-jetson-orin-nano-devkit}},
  note = {Accessed: 2025-03-15}
}

@misc{huang2023infraredinsituirisinspection,
      title={Infra-Red, In-Situ (IRIS) Inspection of Silicon}, 
      author={Andrew 'bunnie' Huang},
      year={2023},
      eprint={2303.07406},
      archivePrefix={arXiv},
      primaryClass={cs.AR},
      url={https://arxiv.org/abs/2303.07406}, 
}

@misc{nv_warp,
  title = {{CUDA} warp size},
  howpublished = {\url{https://docs.nvidia.com/cuda/cuda-c-programming-guide/#hardware-multithreading}},
    note = {Accessed: 2025-03-11}
}

@book{mangard2008power,
  title={Power analysis attacks: Revealing the secrets of smart cards},
  author={Mangard, Stefan and Oswald, Elisabeth and Popp, Thomas},
  volume={31},
  year={2008},
  publisher={Springer Science \& Business Media}
}

@misc{langer_probe_rf,
  key={lprf},
  howpublished = {\url{https://www.langer-emv.de/en/product/rf-passive-30-mhz-up-to-3-ghz/35/rf-b-0-3-3-h-field-probe-mini-30-mhz-up-to-3-ghz/17}},
  note = {Accessed: 2025-01-16}
}

@misc{nv_tensor_core,
  key = {nvtensorcore},
  howpublished = {\url{https://docs.nvidia.com/cuda/cuda-c-programming-guide/\#warp-matrix-functions}},
  note = {Accessed: 2025-01-25}
}

@InProceedings{aes_chosen_input,
author="Moradi, Amir
and Schneider, Tobias",
editor="Standaert, Fran{\c{c}}ois-Xavier
and Oswald, Elisabeth",
title="Improved Side-Channel Analysis Attacks on Xilinx Bitstream Encryption of 5, 6, and 7 Series",
booktitle="Constructive Side-Channel Analysis and Secure Design",
year="2016",
publisher="Springer International Publishing",
address="Cham",
pages="71--87",
isbn="978-3-319-43283-0"
}

@inproceedings{Regazzoni2020:MLHardwareSecurity,
  author       = {Francesco Regazzoni and
                  Shivam Bhasin and
                  Amir Alipour and
                  Ihab Alshaer and
                  Furkan Aydin and
                  Aydin Aysu and
                  Vincent Beroulle and
                  Giorgio Di Natale and
                  Paul D. Franzon and
                  David H{\'{e}}ly and
                  Naofumi Homma and
                  Akira Ito and
                  Dirmanto Jap and
                  Priyank Kashyap and
                  Ilia Polian and
                  Seetal Potluri and
                  Rei Ueno and
                  Elena Ioana Vatajelu and
                  Ville Yli{-}M{\"{a}}yry},
  title        = {Machine Learning and Hardware security: Challenges and Opportunities},
  booktitle    = {{ICCAD}},
  pages        = {141:1--141:6},
  year         = {2020}
}

@inproceedings{Dubey2020:Maskednet,
	title={Maskednet: The first hardware inference engine aiming power side-channel protection},
	author={Dubey, Anuj and Cammarota, Rosario and Aysu, Aydin},
	booktitle={HOST},
	pages={197--208},
	year={2020}
}

@inproceedings{Gongye2023:SCA-DPU,
  title={Side-Channel-Assisted Reverse-Engineering of Encrypted {DNN} Hardware Accelerator {IP} and Attack Surface Exploration},
  author={Gongye, Cheng and Luo, Yukui and Xu, Xiaolin and Fei, Yunsi},
  booktitle={IEEE S\&P},
  year={2024}
}

@inproceedings{Joud2022:PracticalSCADNN,
	title={A Practical Introduction to Side-Channel Extraction of Deep Neural Network Parameters},
	author={Joud, Rapha{\"e}l and Mo{\"e}llic, Pierre-Alain and Ponti{\'e}, Simon and Rigaud, Jean-Baptiste},
	booktitle={CARDIS},
	pages={45--65},
	year={2022},
}

@inproceedings{poster_input,
author = {Batina, Lejla and Bhasin, Shivam and Jap, Dirmanto and Picek, Stjepan},
title = {Poster: Recovering the Input of Neural Networks via Single Shot Side-channel Attacks},
year = {2019},
isbn = {9781450367479},
publisher = {Association for Computing Machinery},
address = {New York, NY, USA},
url = {https://doi.org/10.1145/3319535.3363280},
doi = {10.1145/3319535.3363280},
booktitle = {Proceedings of the 2019 ACM SIGSAC Conference on Computer and Communications Security},
pages = {2657–2659},
numpages = {3},
location = {London, United Kingdom},
series = {CCS '19}
}

@InProceedings{chari_masking,
author="Chari, Suresh
and Jutla, Charanjit S.
and Rao, Josyula R.
and Rohatgi, Pankaj",
editor="Wiener, Michael",
title="Towards Sound Approaches to Counteract Power-Analysis Attacks",
booktitle="Advances in Cryptology --- CRYPTO' 99",
year="1999",
publisher="Springer Berlin Heidelberg",
address="Berlin, Heidelberg",
pages="398--412",
isbn="978-3-540-48405-9"
}

@article{Li2022:PowerAttacksDNN,
	title={Power-based Attacks on Spatial {DNN} Accelerators},
	author={Li, Ge and Tiwari, Mohit and Orshansky, Michael},
	journal={ACM Journal on Emerging Technologies in Computing Systems},
	volume={18},
	number={3},
	pages={1--18},
	year={2022}
}

@inproceedings{gandolfi2001electromagnetic,
  title={Electromagnetic analysis: Concrete results},
  author={Gandolfi, Karine and Mourtel, Christophe and Olivier, Francis},
  booktitle={International workshop on cryptographic hardware and embedded systems},
  pages={251--261},
  year={2001},
  organization={Springer}
}

@INPROCEEDINGS{QS01,
  author       = {Jean{-}Jacques Quisquater and
                  David Samyde},
  title        = {ElectroMagnetic Analysis {(EMA):} Measures and Counter-Measures for
                  Smart Cards},
  booktitle    = {E-smart},
  pages        = {200--210},
  year         = {2001}
}

@inproceedings{schneider2015leakage,
  title={Leakage assessment methodology},
  author={Schneider, Tobias and Moradi, Amir},
  booktitle={CHES},
  pages={495--513},
  year={2015},
}

@inproceedings{Yli2021:ExtractionBNN,
	title={Extraction of binarized neural network architecture and secret parameters using side-channel information},
	author={Yli-M{\"a}yry, Ville and Ito, Akira and Homma, Naofumi and Bhasin, Shivam and Jap, Dirmanto},
	booktitle={ISCAS},
	pages={1--5},
	year={2021}
}

@inproceedings{kocher1996timing,
  title={Timing attacks on implementations of {Diffie}-{Hellman}, {RSA}, {DSS}, and other systems},
  author={Kocher, Paul C.},
  booktitle={CRYPTO},
  pages={104--113},
  year={1996}
}

@inproceedings{kocher1999differential,
  title={Differential power analysis},
  author={Kocher, Paul and Jaffe, Joshua and Jun, Benjamin},
  booktitle={CRYPTO},
  pages={388--397},
  year={1999},
}

@inproceedings{brier2004correlation,
  title={Correlation power analysis with a leakage model},
  author={Brier, Eric and Clavier, Christophe and Olivier, Francis},
  booktitle={CHES},
  pages={16--29},
  year={2004},
}

@inproceedings{kuhn1998soft,
  title={Soft {Tempest}: Hidden data transmission using electromagnetic emanations},
  author={Kuhn, Markus G. and Anderson, Ross J.},
  booktitle={International Workshop on Information Hiding},
  pages={124--142},
  year={1998},
}

@inproceedings{chmielewski2021reverse,
  title={On Reverse Engineering Neural Network Implementation on {GPU}},
  author={Chmielewski, {\L}ukasz and Weissbart, L{\'e}o},
  booktitle={AIHWS},
  pages={96--113},
  year={2021},
}

@inproceedings{DBLP:conf/uss/HorvathL0B24,
  author       = {P{\'{e}}ter Horv{\'{a}}th and
                  Dirk Lauret and
                  Zhuoran Liu and
                  Lejla Batina},
  editor       = {Davide Balzarotti and
                  Wenyuan Xu},
  title        = {SoK: Neural Network Extraction Through Physical Side Channels},
  booktitle    = {33rd {USENIX} Security Symposium, {USENIX} Security 2024, Philadelphia,
                  PA, USA, August 14-16, 2024},
  publisher    = {{USENIX} Association},
  year         = {2024},
  url          = {https://www.usenix.org/conference/usenixsecurity24/presentation/horvath},
  bibsource    = {dblp computer science bibliography, https://dblp.org}
}

@inproceedings{DBLP:conf/iscas/YoshidaKOSF20,
  author    = {Kota Yoshida and
               Takaya Kubota and
               Shunsuke Okura and
               Mitsuru Shiozaki and
               Takeshi Fujino},
  title     = {Model Reverse-Engineering Attack using Correlation Power Analysis
               against Systolic Array Based Neural Network Accelerator},
  booktitle = {{IEEE} International Symposium on Circuits and Systems, {ISCAS} 2020,
               Sevilla, Spain, October 10-21, 2020},
  pages     = {1--5},
  publisher = {{IEEE}},
  year      = {2020},
  url       = {https://doi.org/10.1109/ISCAS45731.2020.9180580},
  doi       = {10.1109/ISCAS45731.2020.9180580},
  bibsource = {dblp computer science bibliography, https://dblp.org}
}

@misc{grattafiori2024llama3herdmodels,
      title={The Llama 3 Herd of Models}, 
      author={LLaMa-Team},
      year={2024},
      eprint={2407.21783},
      archivePrefix={arXiv},
      primaryClass={cs.AI},
      url={https://arxiv.org/abs/2407.21783}, 
}

@misc{carlini2024stealingproductionlanguagemodel,
      title={Stealing Part of a Production Language Model}, 
      author={Nicholas Carlini and Daniel Paleka and Krishnamurthy Dj Dvijotham and Thomas Steinke and Jonathan Hayase and A. Feder Cooper and Katherine Lee and Matthew Jagielski and Milad Nasr and Arthur Conmy and Itay Yona and Eric Wallace and David Rolnick and Florian Tramèr},
      year={2024},
      eprint={2403.06634},
      archivePrefix={arXiv},
      primaryClass={cs.CR},
      url={https://arxiv.org/abs/2403.06634}, 
}

@misc{carlini2020cryptanalyticextractionneuralnetwork,
      title={Cryptanalytic Extraction of Neural Network Models}, 
      author={Nicholas Carlini and Matthew Jagielski and Ilya Mironov},
      year={2020},
      eprint={2003.04884},
      archivePrefix={arXiv},
      primaryClass={cs.LG},
      url={https://arxiv.org/abs/2003.04884}, 
}

@misc{jagielski2020highaccuracyhighfidelity,
      title={High Accuracy and High Fidelity Extraction of Neural Networks}, 
      author={Matthew Jagielski and Nicholas Carlini and David Berthelot and Alex Kurakin and Nicolas Papernot},
      year={2020},
      eprint={1909.01838},
      archivePrefix={arXiv},
      primaryClass={cs.LG},
      url={https://arxiv.org/abs/1909.01838}, 
}

@misc{foerster2024slowsignshighfidelitymodel,
      title={Beyond Slow Signs in High-fidelity Model Extraction}, 
      author={Hanna Foerster and Robert Mullins and Ilia Shumailov and Jamie Hayes},
      year={2024},
      eprint={2406.10011},
      archivePrefix={arXiv},
      primaryClass={cs.LG},
      url={https://arxiv.org/abs/2406.10011}, 
}

@misc{carlini2024polynomialtimecryptanalyticextraction,
      title={Polynomial Time Cryptanalytic Extraction of Deep Neural Networks in the Hard-Label Setting}, 
      author={Nicholas Carlini and Jorge Chávez-Saab and Anna Hambitzer and Francisco Rodríguez-Henríquez and Adi Shamir},
      year={2024},
      eprint={2410.05750},
      archivePrefix={arXiv},
      primaryClass={cs.CR},
      url={https://arxiv.org/abs/2410.05750}, 
}

@INPROCEEDINGS{kariyappa2020adaptive,
  author={Kariyappa, Sanjay and Qureshi, Moinuddin K.},
  booktitle={2020 IEEE/CVF Conference on Computer Vision and Pattern Recognition (CVPR)}, 
  title={Defending Against Model Stealing Attacks With Adaptive Misinformation}, 
  year={2020},
  volume={},
  number={},
  pages={767-775},
  doi={10.1109/CVPR42600.2020.00085}}

@inproceedings{DBLP:conf/asiacrypt/MatherOW14,
  author       = {Luke Mather and
                  Elisabeth Oswald and
                  Carolyn Whitnall},
  editor       = {Palash Sarkar and
                  Tetsu Iwata},
  title        = {Multi-target {DPA} Attacks: Pushing {DPA} Beyond the Limits of a Desktop
                  Computer},
  booktitle    = {Advances in Cryptology - {ASIACRYPT} 2014 - 20th International Conference
                  on the Theory and Application of Cryptology and Information Security,
                  Kaoshiung, Taiwan, R.O.C., December 7-11, 2014. Proceedings, Part
                  {I}},
  series       = {Lecture Notes in Computer Science},
  volume       = {8873},
  pages        = {243--261},
  publisher    = {Springer},
  year         = {2014},
  url          = {https://doi.org/10.1007/978-3-662-45611-8\_13},
  doi          = {10.1007/978-3-662-45611-8\_13},
  bibsource    = {dblp computer science bibliography, https://dblp.org}
}
%\begin{thebibliography}{1}


%\end{thebibliography}
\clearpage
\appendix
\subsection{Far Field Parameter Extraction}\label{sec::app}

\begin{figure}[htbp]
    \centering
    \begin{subfigure}{0.75\columnwidth}
        \centering
        \includegraphics[width=\linewidth]{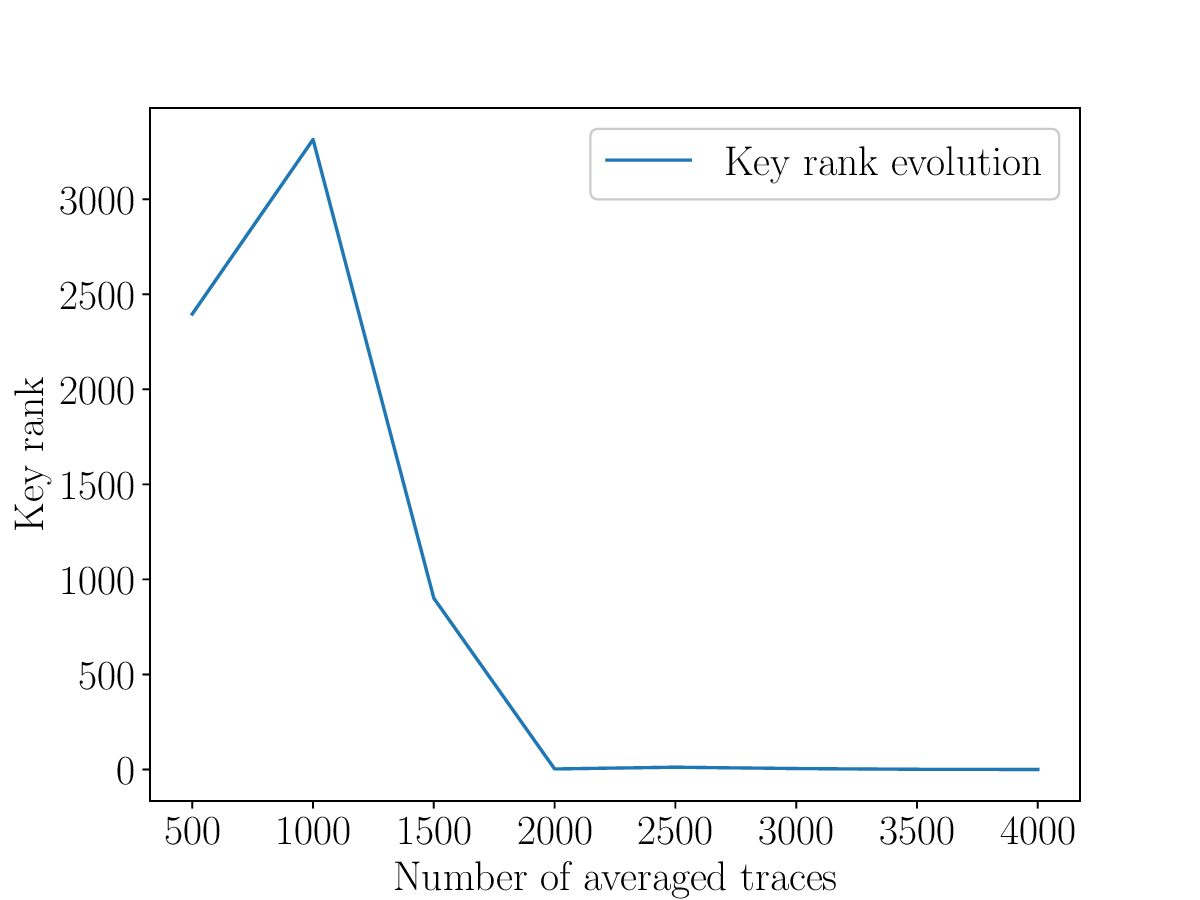}
        \caption{Key rank for the 8th weight in $\mathbf{W_q}$, reaching key rank 0.}
    \end{subfigure}%
    
    \begin{subfigure}{0.75\columnwidth}
        \centering
        \includegraphics[width=\linewidth]{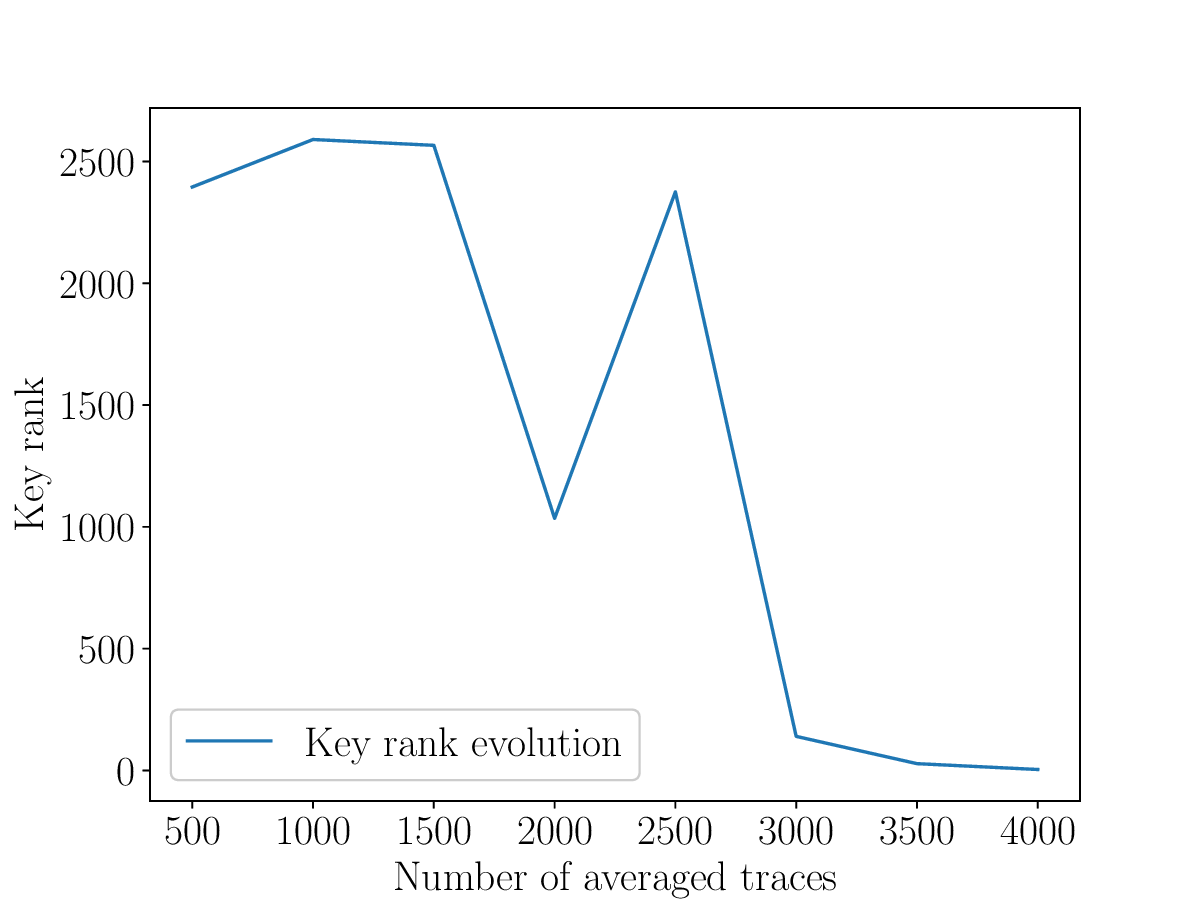}
        \caption{Key rank for the 8th weight in $\mathbf{W_k}$, reaching key rank 4.}
    \end{subfigure}

    \begin{subfigure}{0.75\columnwidth}
        \centering
        \includegraphics[width=\linewidth]{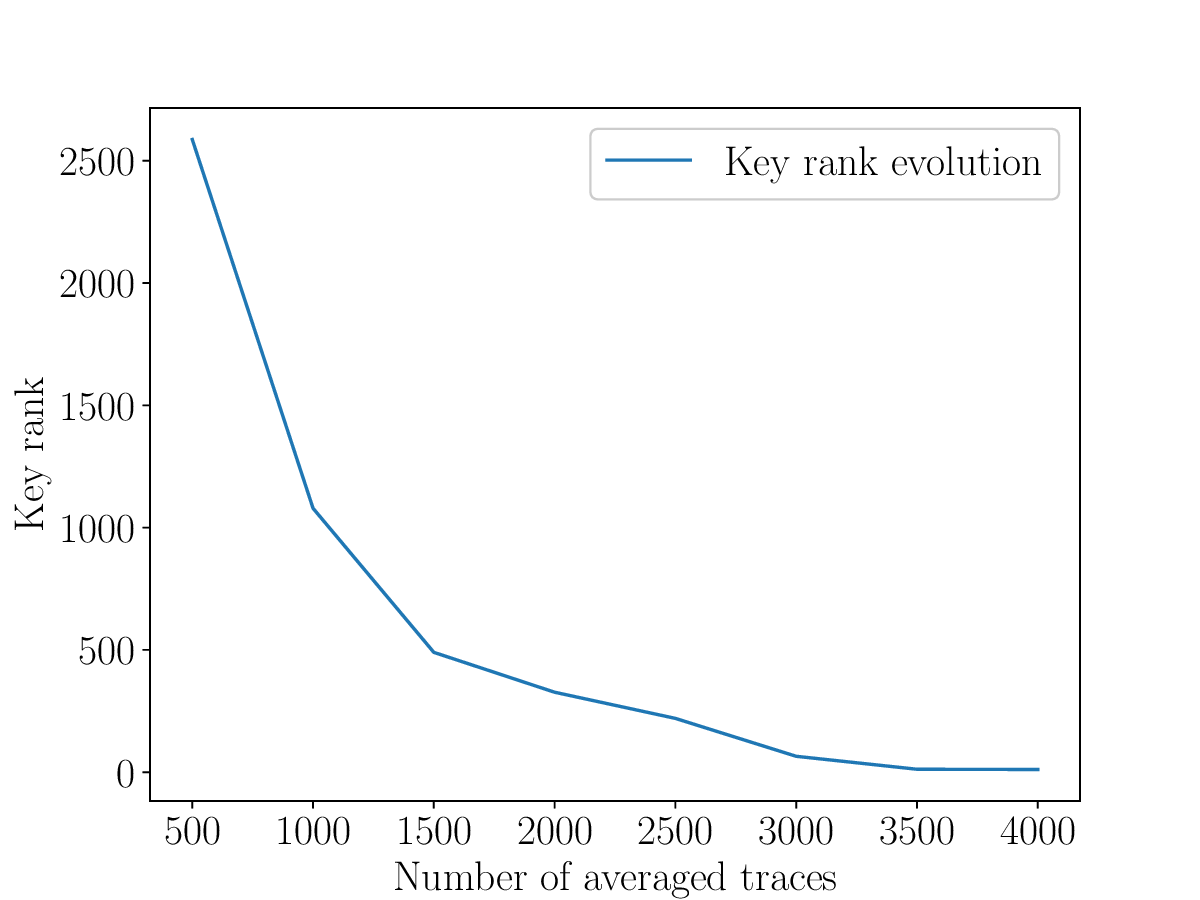}
        \caption{Key rank for the 8th weight in $\mathbf{W_v}$, reaching key rank 11.}
    \end{subfigure}
    
    \caption{Key rankings for different weights in each projection matrix. The number of averaged traces used for CPA is 4000. Averaging is done for traces with the same inputs, with the total of 2 million traces before averaging. In principle, there is weight leakage in far field, but the practical use is limited for the attack.}\label{fig::key_ranks}
\end{figure}

\cref{fig::key_ranks} shows examples of CPA attacks in far field. The examples demonstrate the key rank evolution of CPA for the 8th weight in each of the matrices $\mathbf{W_q}, \mathbf{W_k}, \mathbf{W_v}$. Multiple weights reach low key ranks, proving that, in principle, the developed warp-level models in near field also work in far field.

\end{document}